\documentclass[12pt,preprint2]{aastex} 
 
\newcounter{bbb}
\newcounter{ccc}

\newcommand{\msol}{\hbox{M$_\sun$}} 
\newcommand{\mic}{\hbox{$\mu{\rm m}$}} 
\newcommand{\cms}{\hbox{${\rm cm^{-2}}$}} 
\newcommand{\gcc}{\hbox{${\rm g\,cm^{-3}}$}} 
\newcommand{\cmc}{\hbox{${\rm cm^{-3}}$}}

\newcommand{\umm}{\hbox{${\rm erg}^{-1}\,{\rm cm}^{2}\,{\rm s}\,{\rm \AA}$}} 
\newcommand{\nz}{\hbox{$n_{H}^0$}}

\newcommand{\amed}{$\bar a$}
\newcommand{\amax}{\hbox{$a_{max}$}}
\newcommand{\amin}{\hbox{$a_{min}$}}

\newcommand{\ag}{\hbox{$\gamma$}}
\newcommand{\ad}{\hbox{$\delta$}}
\newcommand{\ap}{\hbox{$\epsilon$}}
\newcommand{\anu}{\hbox{$\alpha_{\nu}$}}

\newcommand{\aox}{\hbox{$\alpha_{OX}$}}
\newcommand{\zp}{\hbox{$z_p$}}

\newcommand{\bla}{\hbox{$\beta_{\lambda}$}}

\newcommand{\zq}{\hbox{$z_q$}}

\newcommand{\HI}{\hbox{H\,{\sc i}}} 
\newcommand{\OIII}{\hbox{O\,{\sc iii}}} 
\newcommand{\NEVIII}{\hbox{Ne\,{\sc viii}}} 
\newcommand{\OVI}{\hbox{O\,{\sc vi}}} 
\newcommand{\CIV}{\hbox{C\,{\sc iv}\,$\lambda\lambda 1549$}} 
\newcommand{\CIII}{\hbox{C\,{\sc iii]}\,$\lambda\lambda 1909$}}

\newcommand{\HE}{\hbox{HE\,2347$-$4342}} 
\newcommand{\HS}{\hbox{HS\,1700+6416}} 
\newcommand{\HStre}{\hbox{HS\,1307+4617}} 
\newcommand{\PGelf}{\hbox{PG\,1148+549}} 
\newcommand{\PGten}{\hbox{PG\,1008+1319}} 
\newcommand{\Pks}{\hbox{Pks\,0232$-$04}}

\newcommand{\Lya}{\hbox{Ly$\alpha$}}

\newcommand{\afuv}{\hbox{$\alpha_{FUV}$}} 
\newcommand{\anuv}{\hbox{$\alpha_{NUV}$}} 
\newcommand{\anuva}{\hbox{$\bar \alpha_{NUV}$}} 
\newcommand{\fdi}{\hbox{$f_{D1}$}} 
\newcommand{\nhz}{\hbox{$n^0_{H}$}} 
\newcommand{\nhf}{\hbox{$n_{H}(z)$}}
 
\newcommand{\nh}{\hbox{$n_{H}$}} 
\newcommand{\Nh}{\hbox{$N_{H}$}} 
\newcommand{\nn}{\hbox{$n^0_{8}$}}
\newcommand{\NN}{\hbox{$N_{20}$}} 
\newcommand{\NNa}{\hbox{$\bar N_{20}$}} 
\newcommand{\mm}{\hbox{$M_{f14}$}} 
\newcommand{\Cut}{\hbox{$C_{\lambda}$}} 
\newcommand{\lao}{\hbox{$\lambda_{obs.}$}} 
\newcommand{\lar}{\hbox{$\lambda_{rest}$}} 

\newcommand{\labr}{\hbox{$\lambda_{brk}$}}

\shorttitle{Intergalactic dust absorption and quasar SEDs} 
\shortauthors{Binette et al.} 
 
\begin{document} 
 
\title{Nanodiamond dust and the far-ultraviolet quasar break}

\date{@@@------> astroph can't handle png figs in pdflatex! {\bf better to download pdf file from http://www.astroscu.unam.mx/binette/qsodust.pdf}}

 

\author{Luc Binette\altaffilmark{1}, Gladis Magris C.\altaffilmark{2}, Yair Krongold\altaffilmark{1},  Christophe Morisset\altaffilmark{1},  Sinhue Haro-Corzo\altaffilmark{1}, Jose Antonio de Diego\altaffilmark{1}, Harald Mutschke\altaffilmark{3}  \\and Anja C. Andersen\altaffilmark{4}
}

\altaffiltext{1}{Instituto de Astronom\'\i a, 
Universidad Nacional Aut\'onoma de 
M\'exico, Apartado Postal 70-264, 04510 M\'exico,  
DF, Mexico; {binette@astroscu.unam.mx.}}
\altaffiltext{2}{Centro de Investigaciones de Astronom\'\i a (CIDA), 
Apartado Postal 264, M\'erida 5101-A, Venezuela; {magris@cida.ve}}
\altaffiltext{3}{Astrophysikalisches Institut und Universit\"ats-Sternwarte (AIU), Schillerg\"a\ss chen 3, D-07745 Jena, Germany; {mutschke@astro.uni-jena.de}}
\altaffiltext{4}{NORDITA, Blegdamsvej 17, 2100 Copenhagen, Denmark; {anja@nordita.dk}}
%
 
 
\begin{abstract} 
We explore the possibility that the steepening observed shortward of
1000\,\AA\ in the energy distribution of quasars may result from
absorption by dust, being either {\it intrinsic} to the quasar
environment or {\it intergalactic}. We find that a dust extinction
curve consisting of nanodiamonds, composed of terrestrial cubic
diamonds or with surface impurities as found in carbonaceous chondrite
meteorites, such as Allende, is successful in reproducing the sharp
break observed.  The {\it intergalactic} dust model is partially
successful in explaining the shape of the composite energy
distribution, but must be discarded in the end, as the amount of
crystalline dust required is unreasonable and would imply an
improbable fine tuning among the dust formation processes. The
alternative {\it intrinsic} dust model requires a mixture of both
cubic diamonds and Allende nanodiamonds and provide a better fit of
the UV break.
The gas column densities implied are of the order $10^{20}\,$\cms,
assuming solar metallicity for carbon and full depletion of carbon
into dust. The absorption only occurs in the ultraviolet and is
totally negligible in the visible.  The minimum dust mass required is
of the order $\sim 0.003 r_{pc}^{2}$\,\msol, where $r_{pc}$ is the
distance in parsec between the dust screen and the continuum source.
The {\it intrinsic} dust model reproduces the flux {\it rise} observed
around 660\,\AA\ in key quasar spectra quite well.  We present indirect evidence
of a shallow continuum break near 670\,\AA\ (18.5\,eV), which would be
intrinsic to the quasar continuum.

\end{abstract}

 
 
 
\section{Introduction} \label{sec:intro} 
 
The spectral energy distribution (SED) of active galactic nuclei (AGN)
contains a significant feature in the optical-ultraviolet (UV) region,
known as ``the big blue bump''.  As for the emission lines in AGN
spectra, it is generally believed that photoionization is  
the excitation mechanism of the  emission lines
superimposed to the continuum.  Photoionization calculations
that reproduce the AGN line ratios and  equivalent widths  favor an ionizing
SED that peaks (in ${\nu}F_{\nu}$) in the extreme-UV
\citep[e.g.][]{mathews,binette88,ferland,korista}. Satellite observations of
distant quasars, however, showed that the big blue bump peaks in
${\nu}F_{\nu}$ around $\ga 1000$\,\AA\ (\lar)\footnote{In the text,
\lar\ and \lao\ will indicate whether the wavelength  refers to the quasar rest-frame or 
the observer-frame, respectively; $\lar = (1+\zq)^{-1} \lao$}
\citep{obrien,zheng,telfer}.  This finding is best illustrated by the
composite quasar SED constructed by \citet{telfer} (hereafter TZ02),
which was obtained by co-adding 332 HST-FOS archived spectra of 184
quasars with redshifts between 0.33 and 3.6.  The composite
(reproduced in Fig.\,\ref{figall} of \S\,\ref{sec:gala}) is
characterized by a mean near-UV index \anu\ of $-0.69$  ($F_{\nu}\propto
\nu^{\anu}$), steepening  to $\simeq
-1.76\pm 0.12$ in the far-UV. The favored interpretation by TZ02, \citet{zheng},  
and \citet{shang} is that the observed continuum steepening is intrinsic to quasars.
Intriguingly, using archived data from FUSE, whose sensitivity window
extends further in the UV, \citet{scott} performed a similar
compilation for `nearby' active galactic nuclei (AGN) with redshifts
$\zq<0.7$ and report the lack of any evidence of a steepening in the
far-UV! Furthermore, the FUSE composite spectrum for nearby AGN is
significantly harder than that of \citet{telfer},
with $\anu \simeq -0.56 ^{+.38}_{-.28}$ in the far-UV. Arguably, since nearby AGN
are on average less luminous, they may possess an intrinsically
different SED. However, a detailed optical-UV study of a subset of the
FUSE sample by \citet{shang} fails to reveal any correlation between the far-UV index
and blackhole mass. Therefore, even though the HST-FOS and FUSE
samples do not represent equivalent AGN populations, the absence of
steepening in the far-UV for the nearby sample cannot be explained  alone by
difference in  AGN populations. A plausible explanation for the
onset and increasing importance of the break in distant AGN can be
provided by intergalactic absorption, since it would scale with
distance. In an earlier paper, \citet{binc} explored the possibility
that the break might be the result of \HI\ scattering by a tenuous
intergalactic component that the authors associated with the warm-hot
intergalactic medium. The models, however, predicted a significant
flux discontinuity in the region 1050--1190\,\AA\ (\lao), which is not
observed in quasar spectra, as shown by FUSE
\citep[e.g.][]{kriss}. Furthermore, the warm-hot intergalactic medium
is too  ionized to produce the amount of
\HI\ absorption needed to reproduce the break. In this Paper, 
we explore an alternative interpretation that is based on a different
opacity vector, namely dust, either intrinsic to the quasar
environment or intergalactic. The vector responsible for the
absorption will consist of grains made of carbon atoms, a major
constituent of the interstellar medium dust, albeit here in
crystalline form (nanodiamonds).  We will assume that the intrinsic
quasar SED consists of a simple powerlaw, and that deviations from the
powerlaw are caused by absorption from crystalline carbon dust, either
as pure cubic diamonds, or of the type observed in carbonaceous
chondrite meteorites (e.g. Allende, Orgueil and Murchison). Many mechanisms have been
proposed to explain the formation of diamond nanocrystallites
(c.f. \S\,\ref{sec:nano}).  About half of them require intense UV 
irradiation. Interestingly, a significant UV flux is present
in the two Herbig Ae/Be objects, for which nanodiamond emission bands
have been confirmed first \citep{van02}.  Processes that form
nanodiamonds by UV irradiation are particularly relevant, since quasars
are UV powerhouses and their environment might lead to physical
conditions that favor the emergence of carbon-based nanocrystallite
grains.

The paper is structured as follows: following the introduction in
\S\,\ref{sec:intro}, we describe in \S\,\ref{sec:cal} the dust models and the algorithm used to
compute the transmission function. The methodology and classification
of the spectra are described in \S\,\ref{sec:appr}. In
\S\,\ref{sec:intr} and \S\,\ref{sec:gala}, we present  the
intrinsic and intergalactic dust absorption models, respectively, and
compare them with the observed spectra. In \S\,\ref{sec:comp} we
decide, which of the two models, is to be preferred, and we discuss a
possible final model. In \S\,\ref{sec:nano} we focus on the formation
and physical properties of nanodiamonds, and follow with the
conclusions in
\S\,\ref{sec:conc}.

\section{Procedure and calculations}\label{sec:cal}

\subsection{Dust extinction curves } \label{sec:ext}

In order to account for the sharp SED break by way of dust absorption,
we looked for an absorption vector that peaks in the far-UV ($\lambda
< 1000$\,\AA), and yet causes negligible absorption at wavelengths
longer than 1200\,\AA. Ideally, as it is the case with the interstellar medium (ISM) dust, the grain
particles should be composed of the most abundant elements. In both
aspects, the crystalline form of carbon is the most appealing
candidate and is the basis of this Paper. We will consider two types
of materials: the terrestrial cubic (pure) diamonds and the
nanodiamonds as found in meteorites.

A comparison of the UV extinction properties of the terrestrial
diamonds and the meteoritic nanodiamonds can be found in
\citet{mutschke}.  The authors separated the nanodiamonds from the
Allende\footnote{Carbonaceous chondrite meteorites are
relatively rare, at a frequency of only $\sim 3.5$\%. The Allende
meteorite who fell on Earth near the town of Allende in the state of
Chihuahua, M\'exico, on February 8th, 1969, is one of the most studied
meteorites of its kind.} meteorite sample and determined their optical
constants. The meteoritic nanodiamonds differ in their optical
properties from the cubic diamond as a result of chemical impurities (e.g. H, N)
and of restructured or unsaturated bonds at their surface. 

Following a standard procedure, we calculated a set of dust extinction
curves. We assumed that the grains are spherical and that a powerlaw
describes the differential distribution of grain sizes, $dn_{gr}(a)/da
= C_{gr} n_H a^{\zeta} $, where $n_{gr}$ is the volume density of
grains, \nh\ that of hydrogen, $a$ the grain radius, $\zeta$ the
powerlaw\footnote{A log-normal distribution is more
appropriate \citep{lewis89} for  describing the nanodiamond size distribution. 
However, because we
intended to explore a much broader  size range than that  found in
Allende, we considered it  more convenient to use a powerlaw  for this
purpose. The reason is that  only one parameter needs to be varied (\amax), while in
the case of the log-normal distribution, we would need to vary  two
parameters simultaneously (the mean size value and the distribution's width).} 
index and $C_{gr}$ the normalization
constant, such that the density of grains becomes normalized to the
abundance of the dust constituents with respect to hydrogen.  We
adopted the tabulated complex refraction indices $n+ik$ of
\citet{mutschke} for the Allende meteorite nanodiamonds, and of
\citet{edwards} for the cubic\footnote{The extinction cross-section for cubic diamonds
were extrapolated in the far-UV, since we could not find published
laboratory values of the refraction indices shortward of 413\,\AA.}
(terrestrial) nanodiamonds.  The Mie theory was used to compute the extinction
cross-section $Q_{ext}(a,\lambda,n,k)$, using a modified version of
the published subroutine BHMIE of \citet{bohren}.  The extinction
cross-section is normalized with respect to the gas density
\nh, using the following integrals:
\begin{displaymath}\label{eq:sig} 
\begin{array}{rl}
\nh \sigma_{\lambda}^{H} &= \nh C_{gr}  {\int_{a_{min}}^{a^{max}} \pi a^{3+\zeta} \,
Q_{ext}(a,\lambda,n,k)  \, d{\rm ln}a}  \\
V_{gr} &= \frac{4}{3} {\int_{a_{min}}^{a^{max}}  \pi a^{4+\zeta}  \, d{\rm ln}a} 
\end{array}
\end{displaymath}
where \amin\ and \amax\ are the minimum and maximum values of the
grain radii considered.  The gas opacity is given by the integration
of $d{\tau}_{\lambda}^{ext}= \nh {\sigma}_{\lambda}^{H} dr$.  Neglecting the
contribution of elements other than carbon to the composition of the
nanodiamonds, we adopt a mean molecular weight of $\mu_{gr} = 12$ for the
grains. The value of the normalization constant $C_{gr}$ is
obtained by solving the following:
\begin{displaymath}
\begin{array}{cc}
Z_C \mu_{gr} m_H = \rho_{gr} C_{gr}  V_{gr} \;  \label{eq:ctec}
\end{array}
\end{displaymath}
where $Z_C$ is the carbon abundance by number with respect to H, $m_H$
the hydrogen atom mass and $\rho_{gr}$ the density of the grain
material. The values adopted for $\rho_{gr}$ are 2.3 \citep{lewis89} and 3.51\,\gcc,
for the Allende and the cubic nanodiamonds, respectively. For the grain size exponent, 
we adopted  $\zeta=-3.5$. As we
neither know the gas metallicity nor the dust-to-gas mass ratio, we
assume all carbon is locked in dust and we adopt the solar value of
$3.63 \times 10^{-4}$ for the C abundance, for the sole purpose of
procuring a convenient normalization.

The continuous lines in Fig.\,\ref{figsigmal}, represent the
extinction curves adopted in this Paper.  The curves labeled D1 and
A1 represent the ``small size regime'' extinction curves for
terrestrial diamonds (red curve) and for meteoritic nanodiamonds 
(blue curve), respectively.  In both extinction models, \amin=3\,\AA\ and
\amax=25\,\AA.  In this regime, decreasing further
\amax\ would not alter the extinction curve.  Our
extinction curve A1 (blue  line) is very similar {\it
in shape} to the mass absorption coefficient curve determined by
\citet{mutschke} (see their Fig. 7 or our scaled  version of it,  the green dotted curve in
 Fig.\,\ref{figsigmal}).  A third dust model, which is also useful, is the Allende curve
labeled A3 (orange curve),  whose grain sizes extend up to
200\,\AA.  The peak cross-sections for the
curves D1, A1 and A3, occur at wavelengths 640, 741 and 787\,\AA,
respectively.  Meteoritic nanodiamonds are known to possess a median
radius \amed\ of $\sim 15$\,\AA.  When increasing \amax\ to a value of $\sim
50$ (75)\,\AA\ for the Allende (terrestrial cubic) nanodiamonds
(respectively), one finds that the peak absorption starts shifting
noticeably to the right and the  absorption profile widens
somewhat. This is illustrated by the two long-dashed line curves in
Fig.\,\ref{figsigmal}, both calculated with \amax=100\,\AA.  The above
mentioned curve A3 further extends the grain size range to
\amax=200\,\AA, which significantly shifts the broad absorption peak
toward longer wavelengths.

As shown in \S\,\ref{sec:intr}, the extinction curves D1 and A1 (or
A3 in \S\,\ref{sec:gala}) can reproduce the wide range of continuum steepening observed in
the far-UV in quasar SEDs.  Both types can induce a sharp absorption
break, although the cubic diamond is more extreme in this respect.
This does not occur with ISM dust extinction. For comparison, 
we plot an ISM dust model from \citet{martin} (with $\zeta=-3.5$) in
Fig.\,\ref{figsigmal}, which consists of silicate and graphite grains of sizes
comprised between $\amax=2500$\,\AA\ and \amin = 50\,\AA\ (black
short-dashed line). It is evident that the customary ISM extinction
curve, while reaching a maximum in the UV, still absorbs significantly
longward of the peak, which gives rise to a shallow change of index,
rather than a sharp break. Grain size  is not the main cause for
such differences in relation to nanodiamonds, but
rather the type of material being considered. To illustrate this, we
show a small grains extinction curve used by \citet{magris} to study
the scattered continuum of Pks\,2152$-$69; it has the same composition
as the ISM model, but the size range is reduced to
\amax = 500\,\AA\ (black dotted line in Fig.\,\ref{figsigmal}). 
The  cross-section redward of the extinction peak remains   
too shallow to reproduce the sharp break of quasars.  \citet{shang}
explored the possibility that the standard ISM or even a SMC-like
extinction curve could reproduce the QSO break. Their conclusions is
that reddening by ISM or SMC-like grains ``is not able to produce the
spectral break seen in the AGN sample, without leaving a clear
signature at longer wavelengths'' (which usually is not seen).
Notice that the cubic diamond curve shows a rather narrow peak at
$\sim 650$\,\AA\ followed by a lower plateau at $\sim 500$\,\AA. These
particularities of the cubic diamond cross-section, together with the
very steep rise shortward of $1000$\,\AA\ are unique features, which
should leave a clear imprint, whenever this material is responsible
for the extinction. The extinction at optical wavelengths due to  nanodiamonds is negligible. 
For instance,  with an opacity of unity at 912\,\AA, the extinction in the $V$ band (\lar) is as 
small as $A_V = 10^{-4}$ and $5 \times 10^{-7}$\,mag in the case of the A1 and D1 
curves, respectively.

The extinction curves D1 and A1 (or A3) as defined above will suffice to
test the dust absorption hypothesis.  Instead of using
optically known materials, one could have treated the
absorption hypothesis as an inverse problem, working out the
extinction curve that succeeds best. Considering that the current study
is mostly exploratory in nature, we consider that it confers a  higher
degree of plausibility to use an empirical curve such as that of the
Allende meteorite, rather than an invented cross-section.

\subsection{Calculation of the transmission curve} \label{sec:tra}


The basic assumption behind the current work is that the 
break observed in the spectra is a manifestation of dust absorption
and is therefore {\it not} an intrinsic feature of the SED.  A key
aspect in evaluating how well the dust absorption hypothesis fares is
to assume that we can extrapolate the powerlaw observed in the near-UV
to the region underlying the break.  Any departure of the observed
spectrum from the extrapolated powerlaw will be modeled as
``absorption''. Only in \S\,\ref{sec:cut} will a broken powerlaw be considered for the
far-UV.  In our notation, the ``true'' or intrinsic quasar SED 
is described by either one of the expressions:

\begin{displaymath}
\begin{array}{rl}
 \,&F_{\nu}^{q}  = \cal{A}\, (\frac{\nu}{\nu_{\rm 0}})^{+\anu} \;,\; {\rm or}\\
 \,&F_{\lambda}^q = \cal{B} \, (\frac{\lambda}{\lambda_{\rm 0}})^{\beta_{\lambda}}= B \, (\frac{\lambda}{\lambda_{\rm 0}})^{-({\rm 2}+\anu)}, 
\end{array}
\end{displaymath}
where $\nu_0$ and $\lambda_0$ (=912\,\AA) are the ionization thresholds
of hydrogen in frequency and in wavelength units, respectively, while
\anu\ and $\beta_{\lambda}$ are the corresponding powerlaw
indices. $\cal A$ and $\cal B$ are normalization constants, one of which is set
to unity according to whether $F_{\nu}^{q}$ (i.e. $\cal A$) or
$F_{\lambda}^{q}$ (i.e. $\cal B$) is plotted (respectively). In keeping with the
tradition in AGN literature, the index that we quote in the text will always
be \anu. In par with the work of TZ02 and
\citet{zheng},  we prefer to   plot $F_{\lambda}(\lambda)$ for most figures. 
The far-UV region beyond the break is better represented using
$F_{\lambda}$ than $F_{\nu}$, as can be appreciated by comparing panels
{\it a} and {\it b} of Fig.\,\ref{figcartoon}.

\subsubsection{Intrinsic dust absorption} \label{sec:def}

The (modeled) transmitted flux in the one-dimension case is given by
$F_{\lambda}^{mod}= T_{\lambda} F_{\lambda}^q$, where $T_{\lambda}$ is
the transmission function, which for a point source is simply the
exponential $e^{-\tau_{\lambda}^{ext}}$.  To compute the opacity $\tau_{\lambda}^{ext} =
N_H \sigma_{\lambda}^{H}$ in the case of  dust  at the
redshift of the quasar, all that is required is to specify the
absorption column \Nh\ and select one, or a combination, of the
extinction curves described in
\S\,\ref{sec:ext}.  

\subsubsection{Intergalactic dust and  the simulation of the composite spectrum} \label{sec:sim}

If the dust is intergalactic, it is necessary to integrate the
transmission along the line-of-sight to the quasar.  Because we also
intend to simulate the process of constructing a composite spectrum
from synthetic quasar SEDs, we developed the following numerical
procedure.  Briefly, the simulation of the composite will consist in
multiplying each synthetic quasar spectra by the appropriate
transmission function and then co-add them in the quasar
rest-frame. The synthetic spectra, before dust absorption, share the
exact same SED, but differ in redshift and in spectral coverage. In the
simulation, we adopted the same set of quasar redshifts as those in
the TZ02 sample as well as the same set of wavelength limits, for the
synthetic spectra, as those characterizing  the  TZ02 archived 
spectra. 
Each synthetic spectrum corresponding to a given quasar at redshift \zq\ is
divided into energy bins, and for each rest-frame bin $\lambda_j$, we
calculate the (modeled) transmitted flux $F_{\lambda_j}^{mod} =
F_{\lambda_j}^q T_{\lambda_j} = F_{\lambda_j}^q e^{-\tau(\lambda_j)}$
making use of  the integrated  opacity along the line-of-sight up to \zq:
\begin{displaymath}
\begin{array}{cc}
\tau(\lambda_j) = {\int_{0}^{z_q} \nhf \;
\sigma^H_{\lambda}\!(\frac{\lambda_j}{1+z})\; \frac{dl}{dz} \; dz}  \label{eq:tau}
\end{array}
\end{displaymath}
where $\sigma^H_{\lambda}$ is the dust extinction cross-section evaluated at
wavelength $\lambda_j/(1+z)$, and \nhf\ the intergalactic dust density
expressed in terms of the hydrogen density. 

For the calculations of distances, $\frac{dl}{dz}$ and baryonic
densities, we assume the concordance $\Lambda$CDM cosmology with
parameters derived from the WMAP experiment \citep{spergel}, that is
$\Omega_{\Lambda} = 0.73$, $\Omega_{M}=0.27$, $h = 0.71$ with $h =
H_0/100$ and a baryonic mass of $\Omega_{bar}h^2=0.0224$ corresponding
to an hydrogen density at zero redshift of $n_{bar}^0= 2.06 \times
10^{-7}$ \cmc.
 
\section{Methodology} \label{sec:appr}

\subsection{The initial database} \label{sec:data}

The spectral database adopted in this work is that of TZ02, which was
kindly  lent to us by R. C. Telfer. It comprises 332 spectra,
mostly HST-FOS\footnote{Note that the TZ02 sample includes 3 HST-STIS
and 6 HST-GHRS spectra.}, of 184 quasars, already reduced and
corrected for Galactic dust extinction. The spectra  furthermore have
been corrected by TZ02 for the presence of Lyman limit absorbers (down
to $\tau > 0.3$) as well as of the \Lya\ absorption valley (caused by
the cumulated absorption from unresolved \Lya\ forest lines).  

\subsection{Near and far-UV spectral indices}\label{sec:ind}

We define the far-UV as the wavelength region shortward of the break
from 300--1000\,\AA, the near-UV as the 1000 to 3200\,\AA\ region
longward of the break, and the optical-UV as the 3200--4200\,\AA\
region.  

Throughout this Paper, we will refer to the powerlaw index
longward of the break as \anuv, and that shortward as \afuv.  We will
assume that the intrinsic SED powerlaw index, \anu, has the same value
in the region of the break as in the near-UV, hence
\anu=\anuv.  Whenever possible, the adopted value for \anuv\ will be the
value that we estimate empirically, using the adjacent near-UV region
of the HST-FOS spectrum. This value is to be preferred over 
published values, which correspond to a forced fit of the combined
optical-UV region. The HST-FOS \anuv\ indices are usually
significantly harder. They are more appropriate for the exercise at
hand, which relies on having a dependable SED description immediately
longward of the break that can be extrapolated one octave shortward,
in the region of the break itself.


\subsection{Pre-analysis of the HST-FOS sample}\label{sec:pre}


Following a preliminary analysis of the TZ02 sample and of the
properties of dust models, we established the following. (I)--
Individual quasar spectra provide  stronger constraints to the
models than a single composite spectrum. The process of co-adding
varied spectra to construct the composite inevitably lead to a loss of
valuable information. Modeling the composite spectrum is probably an
essential exercise, but does not constitute a determinant proof of the
validity of any model. For these reasons, we concentrate here on
fitting {\it individual} quasar SED.  (II)-- Since combined {\it
multigrating} spectra extend over a larger wavelength domain, they
provide stronger constraints for the models than single grating
spectra. For this reason, this work considers only those 106 spectra
of the TZ02 sample that correspond to a combination of two or three
HST-FOS gratings\footnote{Of particular interest are quasars in the
redshift range 0.9--2, for which the spectrum corresponds to a
combination of 3 gratings.  In these, the break is in full view and,
in most cases, there is sufficient wavelength coverage, longward of
the break, to infer the spectral index \anuv\ and, shortward of the
break, to distinguish absorption features that the models must reproduce. }.  
(III)-- In the process of looking for patterns among
the numerous spectral shapes encountered, we found it beneficial to
classify these according to the signs by which dust absorption
apparently manifests itself. The proposed classification is nothing
more than a convenient and simplified characterization of the big blue
bump phenomenology found among the archived HST-FOS spectra.  By no
means it implies that the quasars themselves are intrinsically
different as a result of their spectra belonging to one class or
another.




\subsection{Classification of multigrating spectra into classes A--D}\label{sec:targ}

A physical insight on how dust can alter the continuum shape and
account for the break has led to the classification of the multigrating
spectra into four groups. The three most relevant groups are
qualitatively described in panel $a$ of Fig.\,\ref{figcartoon}. The 4 classes are
defined as follows.

\begin{list}{(\Alph{bbb})\ -}{\usecounter{bbb}}
\item  The spectra that show a continuum steepening  near
$1000$\,\AA\ (\lar) belong to class (A).  The near-UV spectrum is hard
in these spectra and the far-UV shows a moderately steepened
continuum. \PGelf\ ($\zq=0.969$) can be considered the archetype of
this class (see Fig.\,\ref{figpg1148a}).  We tentatively assign the
7 spectra (usually hight redshift quasars), whose near-UV FOS spectrum
is not available longward of 1300\AA, to class (A).  \HS\ ($z=2.722$)
for instance is classified as class (A) (Fig.\,\ref{figcombi}). More
than 60\% of quasars, whose spectra extended sufficiently into the
far-UV to determine \anuv\ belong to class (A) alone. This may
explain, why the spectra of this class individually resemble the TZ02
composite shape, since the composite is after all the result of
averaging spectra that most often than not belong to class (A).


\item The spectra that show a  sharp break near 1000\,\AA, 
followed by an extremely steep continuum drop shortward of the break,
belong to class (B). The near-UV spectrum is hard in these
spectra. PG\,1248+401 ($\zq=1.03$) in Fig.\,\ref{figpg1248} can be
considered the archetype of this class.  Another example is
Pks\,0122$-$00 ($\zq=1.07$) in Fig.\,\ref{figpks0122}.  Objects
in this class are not that common (only 6), though striking\footnote{Ton\,34,
which was reported to have an index of $\afuv = -5.3$ by TZ02, is
another example but only a single grating spectrum exists.} by their
lack of a significant flux in the far-UV.

\item The spectra that show a  continuum  that is already soft longward 
of the break, that is, up to $\ga 1600$\,\AA, belong to class (C). The
soft region of the continuum  now  extends to include  the continuum beneath the
\CIV\ doublet  (or even  down to \CIII\ in some cases).
A representative class (C) spectrum is the quasar 1130+106Y in
Fig.\,\ref{figq1130} ($\zq=0.54$).  In many cases, a single powerlaw
does not fit  the near-UV part well and, in other cases, the index is
very steep ($<-1$) throughout the whole spectrum, as exemplified by
3C279 in Fig.\,\ref{figq1130} (green spectrum).  In the far-UV, these
objects show characteristics of either class (A) or (B), that is,
they are either flat in $F_{\lambda}$ or very steeply declining, as
illustrated by MC\,1146+111 ($\zq=0.863$, blue spectrum), which
exhibits a class (B)-like break. We found 8 objects with the above
characteristics\footnote{Having access only to the near-UV spectrum may suffice for
classifying quasars into class (C) but, in keeping with the above
classification rules, we did not consider nor classify any spectrum that did
not extend down to at least 900\,\AA\ in the far-UV.}. 

\item The high redshift quasars (3 objects) that we could not make sense of, 
belong to class (D). 
They are objects that show an inflection or wide through in the
far-UV.  HE\,1122$-$1649 is one example (see blue spectrum in
Fig.\,\ref{figpks0232b}).  We do not rule out that the troughs could be
associated in some cases to one or more \Lya\ absorption systems.

\end{list}

Only 61 of the available 106 {\it multigrating} spectra extended
sufficiently shortward of the break, that is, down to at least
$900$\,\AA,  to ensure  proper classification.  
Therefore, only this subset of 61 quasars has been analyzed in detail and modeled.
Of these, 44
are class (A), 6 class (B), 8 class (C) and 3 class (D).  We will
mainly focus on class (A) and (B) spectra.  These two groups together
represent 82\% of the classified objects and will suffice for the
purpose of testing the dust absorption hypothesis.  
Many interesting spectra that could not be shown in this Paper will appear elsewhere
\citep[e.g.][b, c]{bi05a}.
As for class (C),
dust appears to be related to some of the observed characteristics of
at least a fraction of them (\S\,\ref{sec:C}) but further work will be
needed to reach definite conclusions. The few objects that form class (D) are
puzzling and will not be modeled with dust absorption in this Paper.

\subsection{Absorption models considered: intergalactic vs intrinsic}\label{sec:cate}

In order to explore how dust absorption might be the real cause of the
observed break, a decision must be made on where the dust is located.
The answer to this question defines two basic\footnote{As the spectra
have already been corrected for Galactic reddening, there is no
reason to be concerned by Galactic dust absorption.}  types of
absorption models: (a) the dust is intrinsic to the environment of the
quasars, and (b) the dust fills the intergalactic space.  In case (a),
the transmission function is derived directly from the extinction
curve in the rest-frame of the quasar, as mentioned in
\S\,\ref{sec:def}. With this category of models, we may reasonably expect
the amount of dust to vary more or less at random from object to
object. In case (b), the dust distribution is intergalactic and fills
large volumes of space.  We therefore expect that such a dust
distribution should be, to a first order, homogeneous, since the dust
becomes a cosmological component unrelated to the quasars. With case (b), 
the models  predict the same transmission for objects of
comparable redshifts, independently of class. The dust density, \nhf,
function of redshift as mentioned in \S\,\ref{sec:sim}, has to be
determined, requiring extra constraints. Intergalactic models imply
enormous amounts of dust, since it is  cosmological. We will first
study the intrinsic dust case and then proceed to the intergalactic
case.

In all figures, when overlaying a dust absorbed  {\it  model} to the rest-frame spectrum
of a particular quasar at redshift \zq, we follow the following
coding: the continuous line part depicts the wavelength region corresponding
to an idealized FOS spectrograph window extending from 1250 to
3600\AA\ (\lao), while the dashed line represents an extension
into the far-UV down to 915\AA\ (\lao), as would be available using
the FUSE satellite.  A dotted line is used outside these two {\it
observer-frame} windows.

\section{The case for intrinsic dust} \label{sec:intr}

The case in favor of intrinsic dust absorption is best made by going through each class in
order of increasing complexity of the dust  model that it requires, that is
in order B, A and C.  

\subsection{Class B spectra}\label{sec:B}

Although class (B) objects are not numerous, they gave us an important
clue on how to disentangle various effects resulting from dust
absorption.  What characterize this class is the very steep drop of
the UV flux shortward of 1000\,\AA\ (\lar). Class (B) objects can
easily be accounted for by simply using the extinction curve D1
consisting of terrestrial cubic nanodiamonds and adjusting as needed
the absorption column\footnote{It is likely that only a  small fraction
of carbon is actually locked  into nanodiamond grains. Supposing we
independently knew the dust-to-gas ratio due to nanodiamonds
$\Delta_{DTG}^C$ and that it was smaller than 0.0031, which
corresponds to depleting all the carbon onto dust, then  the absorption
columns quoted in this paper would have to be multiplied by the factor
$0.0031/\Delta_{DTG}^C$. The galactic ISM dust is characterized by a
much larger $\Delta_{DTG}^{ISM}\simeq 0.009$, since it contains many other
atomic species than C.\label{foot:dtg}}
\Nh. This is illustrated in Fig.\,\ref{figpg1248}, which shows the
spectrum of the archetype class (B) quasar, PG\,1248+401.  The  red
line corresponds to a model using the curve D1 and an  absorption column density
\Nh\ of $3.2 \times 10^{20}$\,\cms\ (hereafter, the notation \NN = 3.2
will be used).  The assumed underlying powerlaw index is
$\anu=\anuv=0.0$, which is the index that best fits the
emission-line-free continuum longward of \Lya. In all our plots, it is
the quasar spectrum that we scale, until an overlap 
with the model is obtained in the near-UV. The resulting spectrum scaling
factor, \mm, is listed in each caption in units of $10^{14}$
\umm. Since we assume the intrinsic SED to be described by a simple
powerlaw (until \S\,\ref{sec:cut}), the plotted models in all figures will
correspond to the function $F_{\lambda}^{mod}= T_{\lambda} \times
(\lambda/912)^{-(2+\anu)}$, and the $y$-axis can be used to infer the
transmission value $T_{\lambda}$ for any value of $\lambda$.

The only free parameter of the above D1 dust model is the column
\NN. The position of the peak in transmission overlaps surprisingly
well with that of the observed spectrum. This is the result of the
very sharp drop in cross-section of curve D1 longward of $\sim
1000$\,\AA\ (see red curve in Fig.\,\ref{figsigmal}), a property unique to cubic
diamonds among dust grains composed of pure carbon. It is important to note
that class (B) objects cannot be accounted for by making use of the
Allende extinction curves A1 or A3, because these are characterized by
a broader absorption peak and  the peak itself is shifted 
toward higher $\lambda$ values. The blue line in
Fig.\,\ref{figpg1248} illustrates the case of using curve A1. The
relative success of the D1 dust model is telling us that if dust were
indeed responsible for the break in class (B) objects, it must 
mostly consist of cubic diamonds. This does not rule out that a small
fraction of the dust may be of the Allende type. This is demonstrated
by the green line, which corresponds to a model with \NN=2.8 and a
linear dust mixture of 85\% of D1 grains and 15\% of A1
grains. Hereafter, we will use the notation of \fdi\ = 0.85 to
represent the fraction of D1 grains, with $1-\fdi$ being the fraction
of A1 grains.

Other class (B) quasars are Pks\,1229$-$02 and Pks\,1424$-$11, which
are quite similar to PG\,1248+401. They  require models with dust columns
of \NN=3.6 and 2.0, respectively, and an extinction curve consisting
totally or mostly of cubic diamond type D1. In Fig.\,\ref{figpks0122},
we show another (B)-type spectrum, quasar Pks\,0122$-$00, which
requires somewhat less dust. For this object, the green line model has
\NN=2.0 and consists of a dust mixture with \fdi=0.8. It
appears that the model with a mixture of dust types D1 and A1 results
in a superior fit than the blue line model with pure cubic diamond extinction
(with \fdi=1.0 and \NN=2.3).

In summary, class (B) spectra require dust that predominantly consists
of cubic diamonds. The particular absorption characteristics of cubic
diamonds fit the observed far-UV steep drop particularly well. Due to
the large dust opacities implied, there subsists little or no far-UV
flux to be observed in these objects shortward of 800\,\AA.  If
photoionization by high energy UV photons is the excitation mechanism
of the  emission lines, it is puzzling to find that the same high
excitation emission lines are observed in class (B) quasars, devoid of
hard UV, as in other objects that do have a hard
continuum \citep[e.g. \HS\ with \afuv=$-0.55$;][]{reimers89}.  A
possibility is that the dust lies outside the Broad Line Region (BLR). In this case, 
the emission line BLR clouds would  be exposed to
an ionizing continuum that is {\it not} absorbed.



\subsection{Class A spectra}\label{sec:A}

\subsubsection{A mixture of the two nanodiamond grain types}\label{sec:pgelf}

Class (A) spectra show a continuum break that is far less
pronounced. The far-UV continuum is often flat\footnote{By `flat' we
mean an approximately horizontal continuum segment in an $F_{\lambda}$
plot, which translates into an index $\bla\simeq 0$, that is,
$\anu\simeq -2$.} in $F_{\lambda}$, with an index \afuv\ of order $-1.7$
shortward of 1000\,\AA.  One may reasonably expect that the absorption
dust columns are simply smaller than in the previous case.  This
is confirmed by models. Unlike class (B) objects, where one species of
dust is clearly favored, class (A) objects generally require an
extinction curve that combines the extinction from Allende
nanodiamonds with that of cubic diamonds, that is, 
nanodiamonds with and without surface impurities (\S\,\ref{sec:ext}). The spectra of
\PGelf\ in Fig.\,\ref{figpg1148a} will serve to illustrate this point. All
 models shown have the same column of \NN=1.05 and the same SED with
\anu=$-$0.2. Either extinction curve D1 or A1 can give rise to a flat 
continuum immediately longward of the break, but the
onset of the  break turns out inappropriate for both the D1 extinction
curve (red line) and the Allende A1 curve (blue line), as seen in Fig.\,\ref{figpg1148a}.
An extinction curve consisting of a mixture with
\fdi=0.6 (60\% D1 and 40\% of A1 grains) on the other hand provides 
quite an acceptable fit to the break (green curve). We found that, by using a proper mixture of the two
grain types, one can fit all class (A) objects.  
Three prominent emission lines stand out above the far-UV
continuum in \PGelf, shortward of 900\,\AA\ (the line identifications shown in the figures
follow those proposed by TZ02). It is interesting to note that
extinction by pure cubic diamonds (red line) gives rise to a narrow
dip near 700\,\AA, a feature not observed in this particular quasar.


\subsubsection{Applying the intrinsic dust model to class (A) spectra}\label{sec:mixt}

A mixture of the two nanodiamond types is very successful in
reproducing the break in all class (A) spectra. More specifically, we can fit all
observed 1000\,\AA\ breaks assuming one of the four following values
of \fdi: 0, 0.3, 0.6, and 1.0. A finer subdivision in most cases is not warranted by
the data, because the spectra very rarely extend
sufficiently in the far-UV for the fit to be sensitive to small
changes in\fdi. In a few cases, only the onset of the break is seen,
and we could not determine with certainty whether the value of 0.3 or
0.6 is more appropriate.  At any rate, the value of \fdi=0.3 appears
to be the most frequent as indicated in the histogram of
Fig.\,\ref{fighistfdi}, but there remains a substantial fraction of
objects that require a different dust mixture.

In Fig.\,\ref{fighistnh}, we present the distribution of gas columns
derived from fitting class (A) and class (B) spectra. The mean \NNa\
value for class (A) is 1.02 with a standard deviation of 0.29.  There
are two high-redshift spectra, for which there is no evidence of dust,
with an upper limit of 0.1, which are \HS\ and \HE\ (they have not
been included in the average).  If we combine class (A) and (B)
spectra into a single group and assume that they are part of the same
population, we derive \NNa=1.20 and a standard deviation of 0.60.  The
distribution shows that the presence of nanodiamond dust is the rule
in quasars rather than the exception. In many quasars, the amount of
dust inferred is comparable. For instance, 39 quasars have a column
$0.6 \le \NN \le 1.4$, which represents 78\% of the 50 class (A+B) spectra.

As for the distribution of \anu\ describing the near-UV continuum, we
obtain a mean value for class (A) of $\anuva= -0.44$ with a dispersion
of 0.21. This average considers only the 22 objects, for which a
reliable estimate of \anuv\ could be determined directly from the HST
spectra. It is significantly harder than the mean value of $-0.69$
reported by TZ02 and the median value of $-0.83\pm{0.04}$ for local
AGN reported by \cite{scott}, presumably, because the softer class
(C) spectra are not included in our average.

\subsubsection{Dust models predict a rise shortward of $\sim 700$\,\AA}\label{sec:rise}

Due to the rapid decrease in the dust extinction cross section  in the
far-UV, shortward of the cross section peak (see
Fig.\,\ref{figsigmal}), an inescapable feature of dust absorption is
that a rise in transmitted flux always occurs in the far-UV, shortward
of $\simeq 700$\,\AA. Of the 7 multigrating spectra that extended down
to 600\,\AA\ and showed evidence of dust absorption, we found evidence
of a sharp flux rise in 4 of them.  This test was not conclusive for
the remaining 3 spectra. The observation of a steep rise in the far UV
is the strongest evidence in favor of dust absorption and will be
presented in more detail.

The first spectrum with a sharp rise is \PGten, which is shown
in Fig.\,\ref{figpg1008}.  We adopt the value of $+0.13$ as near-UV index,
as determined  by \citet{neugebauer} .  The green line
corresponds to an intrinsic dust model with \fdi=0.3 while the gray line
corresponds to \fdi=0.6.  The column in both models is
\NN=1.2. Clearly the model with an extinction dominated by Allende nanodiamonds (red
line) gives a better fit to the break.  Reducing  \fdi\ further
would cause the break 's onset to occur at too long a wavelength (e.g., the \fdi=0
model in Fig.\,\ref{figpg1148a}).

A second example, is \Pks\ of Fig.\,\ref{figpks0232a}.  A fit of the
near-UV continuum favors \anu\ indices in the range $-0.2$ to
$-0.4$. To be definite, we adopt the steeper SED with $-0.4$. We verified that
the same conclusions are reached when using the harder index.  The red line
model, which is more satisfactory, corresponds to pure D1 dust
with\fdi=1.0 and \NN=0.90 while the green line model corresponds to
\fdi=0.8 and \NN=0.93. Due to the predominance of cubic diamonds, a
narrow dip at 650\,\AA\ stands out in models with $\fdi \ga 0.7$.
This dip appears to be saddled by two prominent emission lines, \OIII\
and \NEVIII, both of which are also visible in the composite SED of
TZ02, but not as prominently. Interestingly,
\citet{scott} discuss the nature of a narrow dip seen blueward of the
\NEVIII\ emission in their near-AGN composite. The interpretation they
favor is that of blueshifted absorption by \NEVIII. Another
explanation might be that absorption by cubic nanodiamonds  is
responsible for this feature. Even though the D1 dip is broader, it
might partially be filled by  the \NEVIII\ emission line.

A third example is provided by the much higher redshift quasar,
\HStre, at $z=2.129$, which is plotted in
Fig.\,\ref{fighs1307a}.  There is no HST-FOS spectrum that covers the
near-UV.  Instead, we adopt the value of \anuv=0.0 as inferred from a
spectrum of D. Reimers and reproduced in \citet{koratkar99}. The three
models superimposed to the spectrum in Fig.\,\ref{fighs1307a} have the
same column \NN=1.3 and differ only by their proportion of the D1 and
A1 dust, as follows : gray line
\fdi=0.8, green line \fdi=0.6, and purple line \fdi=0.3.  The green line
model with \fdi=0.6 provides a better fit. It is interesting to note
that the disjoint part (yellow segment) of the GHRS spectrum (taken with
grating G140L) is  not consistent with the far-UV
extrapolation of the models. A solution to  this problem is found in 
\S\,\ref{sec:cut}.  A fourth example is 1623.7+268B,  shown as the dark green spectrum
at the bottom of Fig.\,\ref{figpks0232a}.


To summarize, the intrinsic dust model is phenomenologically very
successful, as it can not only reproduce the break within the dominant
class (A) spectra, but it can also account for the continuum sudden
rise in the far-UV, in the 650--700\,\AA\ region. The extinction curve
that is required to model the far-UV consists of a mixture of the two
nanodiamonds grain types: the cubic diamonds and the Allende type.

\subsection{Class C spectra}\label{sec:C}

This class is defined by the near-UV shape, which is very soft, taking
the appearance of a flat continuum in $F_{\lambda}$ or, in some cases,
of a bump longward of 1200\,\AA.  We have already presented
(\S\,\ref{sec:targ}) the three examples shown in
Fig.\,\ref{figq1130}. The frequent rounded shape of the spectrum from
near to far-UV suggests the possibility that, in some cases at least,
the spectra might be reddened by dust similar to that of the Galactic
ISM. One such case is 1130+106Y (black line spectrum in
Fig.\,\ref{figq1130}), which we tentatively model using the ISM dust
model of \citet{martin} (see black dashed line extinction curve in Fig.\,\ref{figsigmal}). 
It consists of a
mixture of silicate and graphite grains.  The magenta line is a better model in the far-UV. It
combines ISM extinction (80\%) with that of cubic diamonds (20\% of D1
grains by mass). The column is \NN=9.0 and the assumed SED has
\anu=$-0.25$.  Note the significant attenuation of the continuum, which
reaches a factor of 5 near 1000\,\AA. The optical extinction in the
$V$ band due to ISM-like dust is relatively modest, with $A_V=
0.4$\,mag. Not including the contribution from nanodiamonds extinction
would result in a flatter continuum without much of a break, as
illustrated by the yellow line model (pure ISM extinction with \NN=9.0).

Evidence of reddening by ISM dust appears to be present in 
class (C) spectra, but nanodiamond dust is nevertheless required to
explain the break when it is present. A few class (C) spectra show a steep
drop shortward of 1000\,\AA\ as in class (B) spectra. We have not
explored the possibility of a combination of  the three extinction curves 
A1, D1 and ISM.  It is  possible that some class (C) spectra possess a
SED (from near to far-UV) that is intrinsically steeper than in other
classes, with \anu\  in the range $-2$ to $-1.2$.

\section{The case for  intergalactic absorption}\label{sec:gala}

The fact that  the  absorption columns  take on similar values in
the intrinsic case for the numerous class (A)  spectra, invites us to explore
the hypothesis that the dust pervades the intergalactic space instead
of being confined to the environment of each quasar. Following this
hypothesis, the distribution of the dust bears no relation to the
quasars, but is a function  of distance (i.e. $z$). By the same token, we expect the
grains composition to be more uniform in the intergalactic case than
in the intrinsic case. The intergalactic model does not, however,
imply that there cannot be an additional {\it intrinsic} dust
component local to some quasars, as appears to be the case for class
(B) and (C) spectra. On the other hand, the case for intergalactic
dust will be more convincing if a minority of class (A) spectra
require additional absorption above the one provided by the intergalactic
model.
 
The predictive value of the  intergalactic model  resides in the
function chosen to  describe the  dust  density with
redshift. Such a function will not only allow the modeling of the
break in individual quasars but also the simulation of the composite
SED derived by TZ02, following the procedure defined in
\S\,\ref{sec:sim}.

\subsection{Constraining the dust  behavior with redshift} \label{sec:func}

As a working hypothesis, let us assume that the dust is intergalactic and 
 for now consider only  class (A) spectra.
Since the absorption occurs along the line-of-sight to each quasar,
its impact can extend over the whole far-UV domain as a result of the
redshift effect and cosmological expansion.  In a Universe that
evolves and expands, any cosmological quantity such as the density of
the \Lya\ absorbers, the density of of quasars, the star forming rate,
etc, is known to evolve strongly with redshift, that is with time. The
same must apply to the hypothesized intergalactic dust.  There should
exist an epoch (\zp) at which the dust density reaches a peak.  To describe
such a peak, we adopt a parametric form for the dust
density\footnote{The density \nh\ as defined in this work is {\it not}
a co-moving  but rather a local quantity.} \nhf, similar to that
used by
\citet{baldry} to describe the cosmological star formation rate. It
consists of a broken powerlaw joining at redshift \zp:
\begin{eqnarray}
\nhf=
\begin{array}{ll}
\nhz  (1+z)^{\epsilon}   &{\rm for} \;z \le \zp \\
\nhz (1+\zp)^{\epsilon - \gamma}  (1+z)^{\gamma} &{\rm for} \; z>\zp \label{eq:nhf}
\end{array}     
\end{eqnarray}
where \nhz\ is the density at zero redshift, \ap\ and \ag\ are the low and high redshift indices and
\zp\ the intersection of the two truncated powerlaws.  
Hereafter, we will use \nn\ in units of $10 ^{-8}$\,\cmc\ to express the density at zero redshift.

To constrain the parameters describing the function \nhf, we proceeded
as follows.  Since \citet{scott} did not find evidence of a continuum
break in nearby AGN ($\zq<0.7$), this suggests that the peak in
absorption occurred at an earlier epoch rather than  in the local Universe. 
A positive index for \ap, in which absorption increases with look-back
time, will have the effect of reducing the importance of the
1000\,\AA\ break within the local Universe.  Another indication of the
increase in the importance of the break with redshift can be
appreciated in Fig.\,\ref{figalpha}, where we plot
\afuv\ as measured for each quasar by TZ02.  To derive the mean
values represented by the squares, we distributed the measured indices
into 5 redshift bins and then calculated the average \anuva\ within
each bin. (The continuous line simply connects the 5 mean values.)
After trial and error and varying
\ap, we found that similar fits to the break could be obtained, using
any value within the interval $1.5<\ap <3.5$.  To constrain \ap\ more
effectively would have required including the nearby AGN
observations with the FUSE satellite \citep{scott}.  To be definite, we
hereafter adopt the value \ap=+2.

To determine the behavior of \nhf\ at the other redshift end, we can compare
the very high redshift quasar spectra ($z
\ga 2.5$) with those at intermediate redshifts.  As it turns out,
most high-$z$ HST-FOS spectra are single-grating and can't be
used for that purpose.  Fortunately, there exist two quasars with high
S/N and wide spectral coverage that we could analyze in greater
detail, \HS\ and \HE, which are plotted as $F_{\nu}$ in
Fig.\,\ref{figcombi}\footnote{Because of the hardness of \HE\ ($\bla
\sim -3$), using $F_{\nu}^{obs}$ is much more convenient than
$F_{\lambda}^{obs}$.}.  The powerlaw indices for \HS\ and
\HE\  that we adopt are $\anu=-0.55$ \citep[from][]{reimers89} and $+1.70$ 
(inferred from the best model), respectively. The missing parameter
values defining \nhf\ were arrived at using various constraints, as
described below.

Because of the redshift effect, the HST-FOS spectra of both \HS\ and
\HE\ do not cover the  typical break region  at 1000\,\AA\ (\lar).
Viewed from the perspective of intergalactic dust, however, the
absorption break should  have shifted to shorter wavelengths (with $z$), as
demonstrated below in \S\,\ref{sec:gdis}. As a consequence, the
continuum shape's departure from that of a pure powerlaw must be the
result of the hypothesized intergalactic dust, if such a model is to
be of any use. Despite the ragged appearance of both continua in
Fig.\,\ref{figcombi}, caused by the many absorption systems along the
line-of-sight, it is clear that both show a general
curvature\footnote{Note that if we attempt to fit the above curvature
using intrinsic dust, the absorption actually goes the wrong way,
making the transmitted spectrum appear even harder, as illustrated by
the green line model calculated with \NN=0.5 and
\fdi=0.3.\label{foot:combi}} or change of index,  which intergalactic absorption must
be able to explain.  It turns out that such a curvature can be reproduced by
intergalactic dust models using either of the extinction curves, A1 or D1.
To select the appropriate extinction, we required that the dust model
successfully reproduced  the break observed in the lower redshift
spectrum of \PGelf, which is the archetype of class (A).  This second
constraint effectively rules out cubic diamonds as shown by the red
line model of \PGelf\ in Fig.\,\ref{figpg1148b}.  In the case of
intergalactic models with dust curve A1, the break  occurs at somewhat too
short a wavelength (see blue line in Fig.\,\ref{figpg1148b}). To
compensate for the redshift smearing effect, we extended the grain
size range of the Allende nanodiamonds, extending it up to
\amax=200\,\AA. This defines the new extinction curve A3 (see orange curve in
Fig.\,\ref{figsigmal}) used in all our intergalactic
calculations. This extinction curve A3 was found to provide 
an overall better fit to class (A) objects, in the
intergalactic case.

Having selected the optimal extinction curve, strong constraints on \zp\ and
\nhz\ can now be derived by varying these parameters until an acceptable fit  
of the two high-$z$
quasars is found. We found that \ag\ is loosely constrained to
negative values $\la -1.4$. To further constrain the function \nhf\
and \ag, we made use of the composite quasar SED of TZ02 (shown in
Fig.\,\ref{figall}). We simulated this composite by co-adding
synthetic dust-absorbed SEDs of the same redshifts and spectral widths
as those in the actual TZ02 sample, following the procedure described
in \S\,\ref{sec:sim}.  This exercise indicated a preference for
somewhat larger values of \ag\ than the range favored above using the
two high-$z$ quasars. To be definite, we adopted the value of $\ag=-1.5$,
which corresponds to the overlap between the two types of constraints.
The result of combining these different constraints in an iterative
fashion has been that an acceptable fit to the broad curvature of both
high-$z$ quasars of Fig.\,\ref{figcombi} occurs when using the value
$\zp\simeq 0.4$ for the peak dust redshift. The resulting orange line
model, which requires \nn=3.4, now overlays both continua in
Fig.\,\ref{figcombi} as well as the outline of the break in \PGelf\
(Fig.\,\ref{figpg1148b}). In conclusion, the intergalactic dust model
can account for the progressive steepening of the powerlaw index
observed shortward of 500\,\AA\ in \HE.

\subsection{Applying the intergalactic model} \label{sec:appl}

A reasonable expectation of the intergalactic dust model is that it should
apply to all the classes defined in \S\,\ref{sec:targ}.  This is not to say that additional
absorption by a local dust component cannot take place in some quasars.  For instance,
class (B) quasars, although dominated by intrinsic dust, as shown in
\S\,\ref{sec:B}, can also be modeled as the sum 
of intergalactic absorption and absorption by intrinsic D1 dust.  An
example of such complementarity is given by class (B) Pks\,0122$-$00
in Fig.\,\ref{figpks0122}. An orange line model is plotted, which
includes absorption by both intergalactic A3 dust and intrinsic D1
dust of column \NN=1.0.  Except toward the far-UV, this orange line
model (mostly hidden by the foreground red line!) is almost identical to the
previously described green line model (\S\,\ref{sec:B}), which consisted of an
intrinsic dust mixture with \fdi=0.8 and column \NN=2.0. 
A similar comparison can be established 
with the intergalactic orange line model of PG\,1248+401 (Fig.\,\ref{figpg1248}).

When attempting to simulate the composite SED of Fig.\,\ref{figall},
it turns out that \nn\ must be increased, from 3.4 to 4.7. If not,
the model lied significantly above the composite.  Such a model with
\nn\ increased to 4.7 is shown by the purple continuous line in
Fig.\,\ref{figall}. This  simulation though imperfect  is encouraging,
if we consider that our simulation assumed a single powerlaw index
\anu=$-0.6$, while the TZ02 composite sampled widely different energy distributions. 
Furthermore, TZ02 have combined spectra of classes (A)--(D), while the
{\it pure} intergalactic model is intended for class (A) objects.  It
is presumably for that reason that a significant improvement is
obtained, as shown by the orange line model in Fig.\,\ref{figall}, when
one combines intergalactic absorption with intrinsic absorption by
a column as small  as \NN=0.15 of D1 dust.

We have repeated the same exercise as for the intrinsic case
(\S\,\ref{sec:A}) of applying the intergalactic model to all class (A)
spectra.  We found that 20 spectra could be reasonably fitted with the proposed
value \nn=3.4 (the standard model), while 14 spectra 
required either an increase of $\simeq 10$\% in
\nn\ or the addition of an intrinsic column of dust, most frequently
of type D1. Furthermore, 4 and 6 spectra required \nn\ to be {\it
reduced} (approximately) to 2.4 and 3.0, respectively. The database is
therefore not entirely consistent with the expectation of an
homogeneous distribution of intergalactic dust.

\section{Merits of intergalactic vs intrinsic models} \label{sec:comp}

Which model is to be preferred? We will compare the merits and
problems of each type of model and, in conclusion of this section, we
will present a final model for the 1000\,\AA\ break of quasars.

\subsection{The intrinsic dust hypothesis}\label{sec:idis}

We have shown that {\it intrinsic} dust models could account not only
for the break, but also for the flux rise at shorter wavelengths
(e.g. Figs.\,\ref{figpg1008}--\ref{fighs1307a}).  Overall, the
intrinsic model is  extremely successful across the whole class (A) and class (B)
samples. 
The  fit to the far-UV rise in \HStre, on the other hand, is not entirely satisfactory. 

Is it possible to simulate the TZ02 composite assuming only intrinsic
dust? One difficulty is that the intrinsic model is mute  about
how other parameters like \NN, \fdi\ or \anu\ might vary with
increasing $z$.  However, since each class (A) spectrum can be
fitted quite well by varying \fdi\ and \NN, which is an approach that we
consider superior to that of simply fitting the TZ02 composite, it can
then be argued that not being successful in the simulation of the
composite is of secondary importance. Although we consider this to be
true, we nevertheless attempted to simulate the composite, because it
reveals real trends in quasar SEDs. The fact that the TZ02
composite remains very soft at very short wavelengths instead of showing a far-UV rise,
must be explained somehow.

The silver dashed line in Fig.\,\ref{figall} illustrates our initial attempt
to simulate\footnote{In a situation in which none of the parameters
defining the intrinsic model varies with redshift (as for the silver dashed
line model in Fig.\,\ref{figall}), all the absorbed SEDs are identical
and there is no need to co-add the spectra in order to simulate the
composite.} the composite SED, assuming  \anu=$-0.6$ and keeping  all the input
parameters constant with $z$.  The column is \NN=1.0 and \fdi=0.3.
The far-UV flux is obviously predicted too strong. This happens,
because the extinction cross-section falls off too rapidly at very
short wavelengths, and a steep rise in $F_{\lambda}$ becomes
unavoidable shortward of 600\,\AA. The simulated composite simply
tends toward the slope given by the index \anu. If we vary the column
with redshift, by defining a function $\NN(z)$, we obtain the absurd
result that, in order for the simulated composite to overlap the TZ02
composite, the dust column would have to increase sharply with
redshift. This is not only an  ad\,hoc dust behavior, it is
also contradictory to the absence of any increase with \zq\ of the 
columns determined in \S\,\ref{sec:A}. 
In addition, it is  at odds with  the lack of  absorption  in
the two high-$z$ spectra of \HS\ and \HE, for which we determined
absorption upper limits of $\NN \le 0.1$.  
Attempts to model the curvature in these two spectra with intrinsic dust result in
absorption features at the wrong end of the spectrum. In effect, local
dust makes both spectra appear even harder than they already are, as
illustrated by the green line model in Fig.\,\ref{figcombi} calculated
with \NN=0.5 and \fdi=0.3.

\subsection{The intergalactic dust hypothesis}\label{sec:gdis}

The few shortcomings mentioned above for the  intrinsic case disappear with
{\it intergalactic} dust. By construction, the continuum of the
two high-redshift quasars, \HS\ and \HE, can be reproduced.  On the
other hand, the break can be fitted only for a qualified majority of class (A)
spectra, while the other spectra  usually requiring intrinsic dust to be
added to the model. The TZ02 composite can be reproduced, albeit with a density
\nn\ increased by 40\% (for which we have no satisfactory explanation
to propose).  The far-UV index \afuv, when evaluated at the fixed
wavelength of 800\,\AA, exhibits the correct trend with redshift, as
shown by the long-dashed line in Fig.\,\ref{figalpha}. One may argue
that the amount of dust implied by the intergalactic model is
excessive if not plainly unreasonable, but it is not an impossible amount.
The fraction of the baryonic mass that the value
\nn=3.4 corresponds to is 17\% (see \S\,\ref{sec:cut}) , assuming that the mean cosmic carbon
metallicity is about solar and that the dust is intergalactic, because it
was expelled from galaxies by radiation pressure \citep{ferrara91} or
through supernovae of type II.

The intergalactic model, on the other hand, makes stringent
predictions about how the break ought to shift (and soften) with
increasing redshift.  This is shown in Fig.\,\ref{figtrans}, in which the
transmission function is plotted at representative \zq\ values. The
continuous part of each $T_{\lambda}$ curve corresponds to the
fiducial spectrograph window of 1250--3600\,\AA\ (\lao) [see
\S\,\ref{sec:cate}] and shows what part of the break is visible at a given redshift \zq. 
Notice that when the redshift exceeds values of $\simeq 1.5$, the
break is markedly shifted toward shorter wavelengths.  Of the three
spectra presented in \S\,\ref{sec:rise}, which showed a clear flux
rise in the far-UV, only one is of sufficiently high redshift to test this,
\HStre\ with $z=2.129$.  It's spectrum is shown again in 
Fig.\,\ref{fighs1307b} and can be compared with the pure intergalactic
model with \nn=3.4, which is represented by the brown dashed line. The
SED is the same as earlier, that is \anu=0.0. 

The gradual break (or curvature) seen in the  brown dashed line  model in
Fig.\,\ref{fighs1307b} not only occurs at very short wavelengths
($\sim 550\,$\AA), but is extremely shallow.
Obviously, in order to fit the sharp break characterizing the \HStre\
spectrum, additional intrinsic absorption must be considered.  Such a
model is represented by the orange line, which is a model that
combines intergalactic with intrinsic dust. The local dust column is
\NN=0.8 with a dust composition \fdi=0.6. 
The fit to the observed flux rise is surprisingly good, much better
even than with the pure intrinsic case represented by the green line in
the previous Fig.\,\ref{fighs1307a} of the same quasar.  Even more
suggestive is the disjoint spectrum obtained with GHRS grating G140L
(yellow spectrum in Fig.\,\ref{fighs1307b}), which despite its lower
S/N appears to prolong the far-UV rise of the multigrating spectrum
(black line). The intergalactic model is marginally consistent with
the continuum level set by this spectrum segment, in contrast with the
pure intrinsic model, which rises too steeply (see green line in
Fig.\,\ref{fighs1307a}).  The green dashed line in Fig.\,\ref{fighs1307b}
represents the contribution of intrinsic dust absorption that is
present in the  orange line mixed model. In summary,
even though intrinsic dust is the main contributor to the sharp break observed
in \HStre, the signature at the shortest wavelengths expected in
high-$z$ spectra as a result of intergalactic dust appears to be
independently confirmed in this quasar.

Is the intergalactic hypothesis vindicated? It turns out not to be the
case. In effect, a rather poor fit is provided for the two other
quasar spectra that showed a far-UV flux rise, \PGten\ and \Pks.  As
is the case for \HStre, these two quasars require additional intrinsic absorption, with dust
columns (\NN) of 0.5 and 0.25 and dust mixtures (\fdi) of 0.0 and
1.0, respectively.  However, even when combining this additional
absorption with intergalactic dust, the far-UV rise cannot
be reproduced at all, as shown by the corresponding orange lines in
the Fig.\,\ref{figpg1008} of \PGten\ and in the new Fig.\,\ref{figpks0232b} of \Pks. The
far-UV continuum level is predicted too low in the \Pks\ model and, in both
figures, the flux rise (orange lines) occurs at too short a
wavelength.

\subsection{Evidence of a higher energy break?} \label{sec:cut}


As indicated above, the discrepancy of the intergalactic model for reproducing the far-UV rise 
in \PGten\ and \Pks\ could not be resolved.
This inadequacy of the model is sufficiently significant to reject
the intergalactic dust hypothesis at the assumed
density\footnote{If we adopt an intergalactic model that uses
significantly less dust than \nn=3.4, it would be at the cost of
having more intrinsic dust present and this in a larger fraction of
class (A) spectra, if not the majority of them.  Hence, such a model
would not contribute in an essential way in explaining the break and
would have to be discarded on account of Ockham's razor principle.}.
Moreover, in more than one aspect, the intergalactic model is implausible.
Assuming that the mean metallicity of galactic matter (stars and interstellar gas) at current epochs 
is near solar, as derived by \citet[][b]{calura04a}, then the fraction of cosmic carbon required is
$\nhz/n_{bar}^0=3.4 \times 10^{-8}/2.06 \times 10^{-7}=0.17$.
Apart from the unreasonable fraction of cosmic carbon
that must exist in crystalline form (17\%), it would
require an improbable fine tuning so that this dust would not be
accompanied by larger amounts of the more common flavors (silicates,
graphites, PAHs, ...)  as one would expect if it was formed in supernovae and later
expelled in the intergalactic space. The analysis of cosmological
supernovae has ruled out the existence of large
quantities\footnote{We have calculated the extinction that ISM dust
would produce if it followed the same intergalactic
distribution as the nanodiamonds (\S\,\ref{sec:func}) with \nn=3.4, that is
in amounts that correspond to 17\% of the Galactic dust-to-gas ratio. We
find that for an object at $z=0.5$, the  selective extinction is
$E_{B-V}=0.022$ (\lao),  much in excess  of the value of 0.002 inferred from cosmological supernovae
by \citet{perl99}.}  of intergalactic dust of `normal' composition \citep{perl99}.


Ruling out the intergalactic model in favor of the intrinsic model 
has an additional interesting consequence. 
The curvature present in the spectra of \HS\ and
\HE\ must now be considered an intrinsic feature of the energy
distribution in these two quasars, rather than the manifestation of
intergalactic absorption.

We are left with the intrinsic dust hypothesis to  account for the
1000\,\AA\ break in quasars.  What could  be missing
from this model so that the  few remaining problems  it has
could be resolved? We will
hypothesize that the shallow rollover observed in the far-UV in
\HS\ and \HE\ is a manifestation of a universal high energy cut-off
present in all quasar SEDs.  To test this,  we first model the cut-off
in a way that does not depend on  the \anuv\ of the  underlying SED. 
This is achieved by singling out the transmission curve calculated at
$\zq= 2.8$ (corresponding to the average redshift value of the two
high-$z$ quasars) and considering it to be a valid description of the
continuum softening that might also apply to the other quasars.  We obtained
a parametric fit of this transmission curve at that redshift, using
the following expression:
\begin{eqnarray} \label{eq:cut}
\begin{array}{cc}
\Cut = \left(1+ \left[{\lambda}/{\lambda_{brk}}\right]^{f \delta} \right)^{-f^{-1}}
\end{array}
\end{eqnarray}
where \ad\ is the powerlaw index change and $f$ a form factor, 
that may vary from object to object. In essence, when a
quasar SED is multiplied by \Cut\ and $\ad<0$, a shallow steepening takes
place at \labr, which increments the underlying powerlaw index by \ad.  The
sharpness of the cut-off is set by the form factor $f\ge 1$.  Hereafter, we will
assume that the true intrinsic quasar continuum is  given by
$\Cut \times F_{\lambda}^q$, where $F_{\lambda}^q$ is the quasar powerlaw
that we have been using so far and
\anu\ the index set by the near-UV continuum.  To illustrate the
role of the form factor, let us take the 
 spectrum (blue spectrum in Fig.\,\ref{figpks0232b}) of class (D) quasar
HE\,1122$-$1649 as an example, given that it shows a clear far-UV
cut-off. Longward of the cut-off, the powerlaw index is $\anu=-0.6$,
reaching $-2.6$ shortward of the break situated at $\simeq 670$\,\AA.
Hence the parametric function \Cut\ is characterized by $\ad=-2.0$ and
\labr=670\,\AA.  The gray  and magenta lines in
Fig.\,\ref{figpks0232b} correspond to form factors of $f=2.8$ and 10, respectively.

Turning our attention to the $T_{\lambda}$ curve at a redshift \zq=2.8 in
Fig.\,{\ref{figtrans}, the parametric fit using Eqn.\,\ref{eq:cut}
yields the values $\delta=-1.6$, $f=2.8$ and \labr=670\,\AA.  It is
shown by the thick gray dashed line.  This cut-off can now be
applied, not only to \HS\ and \HE, but to any other quasar as well.
It  will result in a new SED that is characterized by 
the same near-UV powerlaw as before, but with a shallow far-UV cut-off centered on
670\,\AA, that is, at 18.5\,eV.

If we incorporate the above far-UV rollover into our previous quasar
SED of index $\anu=-0.6$, we obtain the distribution represented by
the black dotted line in Fig.\,\ref{figallb}. Assuming this modified SED, a
remarkable improvement in the simulated composite is now obtained, as
shown by the dark gray continuous line in Fig.\,\ref{figallb}. (The silver dashed line model
represents the earlier case without the far-UV intrinsic cut-off).  Further
improvements of the simulated composite can be obtained by varying
some of the input parameters with redshift. The study of these more complex
models, however, exceed the scope of this paper and will be reported
elsewhere.

It is important to emphasize that the intrinsic model with a shallow
cut-off fits markedly better two of the three quasar spectra that
showed a far-UV rise. For instance, the earlier problem encountered
(\S\,\ref{sec:rise}) in fitting the far-UV rise in \HStre\ (green line
in Fig.\,\ref{fighs1307a}) has now disappeared, as shown by the cyan
line model in Fig.\,\ref{fighs1307b}, which assumes the SED with the
above far-UV rollover.  An improvement of the fit to the far-UV rise
in \Pks\ is also obtained, as shown by the cyan line model in
Fig.\,\ref{figpks0232b}. In the case of \PGten, it makes little difference
whether the high energy cut-off is there or  not (compare the cyan and green line models
in Fig.\,\ref{figpg1008}).

In summary, the whole database as represented by the TZ02 composite, as
well as the individual spectra showing the far-UV rise,  are both 
consistent with the presence of an 18.5\,eV\ cut-off in quasars. This
proposed new cut-off is being masked in the vast majority of quasars with redshift $\la
2.5$ by the more prominent 1000\,\AA\ break, which we believe is entirely due to 
nanodiamond dust absorption.

\section{Nanodiamonds: the infrared-UV connection} \label{sec:nano}

Nanodiamonds are, to date, the most abundant presolar grains, both in
mass and numbers, that have been extracted from primitive carbonaceous
meteorites \citep[][and references therein]{mutschke}, but their
detection in the ISM has been elusive. Diamond crystallite emission bands in the
3.3--3.6\,\mic\ region due to surface C--H stretching modes of
hydrogenated nanodiamonds have been established with confidence for a
few Herbig Ae/Be objects and one carbon-rich post-AGB star HR\,4049
\citep[][]{guillois99,van02,acke04}.  \citet{van02} presented  a detailed analysis of
the ISO-SWS spectra of the two Herbig objects, HD\,97048, and
Elias\,I, as well as of the post-AGB HR\,4049. They applied a physical
model to the emission profile of the 3.53\,\mic\ band and inferred a
temperature of 950\,K and 1000\,K for HD\,97048 and Elias\,I,
respectively. Assuming radiative equilibrium between photoheating and
far-infrared cooling for the grains, the authors could estimate the UV
radiation flux impinging the diamonds in these three objects. The
diameter range they inferred for the crystallite diamonds is $2a \sim
10$--100\,\AA.  Interestingly, the multiwavelength data for both
HD\,97048, and Elias\,I, as well as for the post-AGB star HR\,4049,
indicate that the 3.53\,\mic\ emission takes place within a disk-like
structure. The distance between the star and the emission region is
$\le 9$ and $\le 22$\,AU, in HD\,97048 and Elias\,I, respectively.
The formation site that these authors favor for the crystallite
diamonds is in\,situ formation within the disk rather than within the
ISM or via ejection from stars.  In the field of AGN, sub-arcsecond
VLT observations by \citet{rouan04} using NAOS+CONICA revealed
wavelike structures in the mid-infrared, which  the authors propose might 
be due to emission by nanodiamonds at a temperature close to sublimation.


To explain the predominance of nanodiamond grains in primitive
meteorites, several formation mechanisms have been proposed, such as:
(a) chemical vapor deposition from stellar outflows \citep{lewis87},
(b) impact shock metamorphism driven by supernovae \citep{tielens87},
(c) energetic ion bombardment by a supernova \citep{daulton95}, (d) UV
annealing of carbonaceous grains
\citep{nuth92}, (e) nucleation in  organic ice mixtures by UV photolysis
\citep{kouchi05}, and (f)  chemical conversion of PAH clusters to
nanodiamonds in the presence of UV radiation \citep{duley01}.  It is
interesting to note that the last three processes involve UV radiation.  The
above post-AGB and the two Herbig Ae/Be stars emit UV radiation
\citep{van02}, a fact which is possibly 
related to the formation of the observed nanodiamonds.

Could a similar formation process operate within the UV-intense
environment of quasars?  The indication that cubic diamonds dominate
in class (B) quasars might be related to an evolutionary sequence of
the grains.  A possible scenario might be the following. Via the
process of dehydrogenation of PAH clusters by quasar UV radiation,
hydrogenated nanodiamonds form, with optical properties similar to the
Allende type. If the UV radiation heats up the nanodiamonds beyond
1300\,K, a process of surface dehydrogenation begins, which may   
cause the grain optical properties to become  more similar to that of cubic
diamonds. Finally, the disappearance of H--stretch cooling may
result in a runaway heating, followed by graphitization
and eventually to sublimation of the grains.


To confirm the existence of nanodiamond grains in AGN, one could attempt to detect 
the far-infrared emission bands caused by hydrogenated nanodiamonds 
\citep{van02,jones04a}. However, AGN are  intrinsically very strong
far-infrared emitters, and the signature of any narrow emission band
will certainly be diluted. For instance, the dust silicate feature at
9.7\,\mic\ predicted by calculations \citep{laor93} is not observed as
often as expected in AGN. The UV radiation absorbed by nanodiamonds
represent at most 10\% of the energy integrated over the whole SED.
Let us  assume the AGN unification picture with a bi-cone opening angle of
45\degr. If the nanoparticles are located outside the BLR, then a
fraction of only $0.10 \times (1-\cos \case{45\degr}{2}) \la 0.01$ of
the quasar bolometric luminosity will be reprocessed into far-infrared
emission by nanodiamonds. Assuming a uniform covering factor of unity
within the radiation bi-cone, the minimum dust mass required by the
intrinsic model is given by $0.044 \NN\ r_{pc}^2 \times (1-\cos
\case{45\degr}{2}) $\,\msol, where $r_{pc}$ is the distance in parsecs
separating the dust screen from the central UV source. For an arbitrary
distance of one parsec and \NN\ of unity, the dust mass implied is
$0.0033$\,\msol, a value independent of the assumed dust-to-gas ratio
(see footnote\,\ref{foot:dtg}).

To the extent that Allende-type nanodiamonds might be a candidate
carrier \citep[see][]{jones04b} for the Extended Red Emission
observed in nebulae between 5400 and 9500\,\AA, 
it is conceivable that a fraction of the UV flux absorbed by
dust might be re-emitted by photoluminescence.

\section{Conclusions} \label{sec:conc}
 
We have presented evidence that indicates that dust absorption by
nanodiamonds is successful in reproducing the 1000\,\AA\ break as
well as the far-UV rise seen at shorter wavelengths. Could the
agreement between the intrinsic dust models and the spectra be simply the result of a
coincidence between the break location and the far-UV extinction
properties of nanodiamonds? To rule out such a possibility will require
that an independent confirmation of the presence of nanodiamonds be
found. Detection of grain emission in the far-infrared, at 3.43
and 3.53\,\mic, is one possibility, although this emission mechanism
works only with surface-hydrogenated grains (see \S\,\ref{sec:nano}).
Another possible route would be to  observe selected quasars in
order to extend the UV coverage in objects for which only the onset of
the break is seen so far. The idea would be to look for a
confirmation of a flux rise shortward of 700\,\AA\ in as many
quasars as possible.  This would require high quality observations
using a satellite with far-UV sensitivity.  One possibility might be
the R=1000 spectrometer on-board the projected World Space Observatory
satellite, which is expected to offer a sensitivity window covering the range
1100 to 3500\AA\ (\lao) \citep{barstow}.

We have ruled out that the dust causing the 1000\,\AA\ break  
is predominantly {\it intergalactic} on the 
basis that it is not required per\,se and that the
far-UV rise could not be  modeled satisfactorily using intergalactic dust.  Furthermore, the amount of
crystalline carbon that is needed turns out to be improbably large.
Since intergalactic dust is not responsible for the continuum rollover observed in
\HS, \HE,\ at $z\simeq 2.8$, we have proposed  that this feature is a manifestation 
of a higher energy break, near 18.5\,eV, which is presumably
intrinsic. Including the same break in the other quasars SED markedly
improve the simulation of the composite as well as the detailed
modeling of the far-UV rise in \PGten\ and \Pks.  At any rate, such a
break is bound to take place somewhere in the far-UV so that the quasar
SED  connects smoothly with the soft-X rays.  In effect, the optical-X-ray index,
\aox, which relates the monochromatic continuum flux at 2500\,\AA\
to that at 2\,keV, is characterized by values in the range 1.3 to 1.6
(equivalent to an \anu\ between $-1.3$ and $-1.6$).  Given that our mean
\anuv\ index for class (A) is much harder, with $-0.44$, a continuum
turnover must take place somewhere in the far-UV. Our results suggest
that such turnover occurs at 18.5\,eV. We are currently in the process
of studying how the far-UV and the soft X-rays may join together
\citep{sinhue}.

Intrinsic dust models require gas columns of the order $10^{20}\,$\cms,
assuming solar C abundance and full depletion onto nanodiamond grains.  
For comparison, in the solar neighborhood, a $V$-band extinction of a
tenth of a magnitude by ISM dust corresponds to a gas column of $1.9
\times 10^{20}\,$\cms\ \citep{whittet}. 
Larger columns (but same dust masses) would be implied for our models, 
if we assumed  smaller dust-to-gas ratios.  
To the extent that the nanodiamond dust lies in the vicinity of the
AGN, the minimum dust mass required by intrinsic models is small $\simeq
0.003 \NN r^2_{pc}$\,\msol\  (see \S\,\ref{sec:nano}). This value is
independent of the assumed dust-to-gas ratio, because the absorption gas
columns scale inversely with it (see footnote\,\ref{foot:dtg}).

We have shown evidence that ISM-like dust might be playing a role in
explaining the continuum appearance of a fraction of quasars, the
so-called class (C) quasars. To confirm this suggestion will require 
more complex models than presented here, in which more extinction
components might have to  be contemplated (ISM, SMC, nanodiamonds, SiC, ...).
An alternative is that class (C) quasars emit with an intrinsically much
softer SED.

How does fare the intrinsic dust hypothesis  in relation to the result
of \citet{scott}, in which the break is apparently absent in local
Universe AGN? An interesting scenario comes to mind.  An inspection of
the 45 multigrating spectra, which could not be classified, because
their spectra did not extend down to 900\,\AA\ (\lar, see
\S\,\ref{sec:targ}), reveals that the softer class (C) spectra are
more frequent at lower redshifts. Interestingly, a fraction of class (C) quasars do not
show any break (see spectrum of 3C279 in Fig.\,\ref{figq1130}). A
possibility might be that there is a secular evolution of the dust
properties, with nanodiamonds being absent in quasars with $\zq \ga 2.5$,
later becoming predominant at $\zq <2$, and, finally, being
progressively replaced by ISM-like dust for $\zq\la 0.7$.

Interpreting the 1000\,\AA\ break in terms of dust absorption may contribute
to resolve the following  issues in the AGN field:
\begin{list}{(\alph{ccc})\ -}{\usecounter{ccc}}

\item The continuum rise in the 650--700\,\AA\ region, seen in a few individual quasar 
spectra, is not predicted by any accretion disk model or any other
continuum emission model known to the authors. Such a rise 
on the other hand is expected with nanodiamond dust absorption.

\item If the  1000\,\AA\ break is  intrinsic to the ionizing continuum of quasars,  
the mean ionizing photon energy turns out rather small, making it
difficult to account for the observed luminosities of the high
excitation lines \citep[e.g.][]{korista}.  With the alternative interpretation of  dust
absorption, the SED turnover is pushed to higher energies and the
break is an artifact of line-of-sight dust absorption.

\item The puzzling fact that the high excitation emission lines in UV deficient 
quasars (e.g. class (B) quasars such as PG\,1248+401 in
Fig.\,\ref{figpg1248}) are comparatively as luminous as in other
quasars. This is easily explained by dust absorption provided the dust
lies outside the BLR or under any geometry, in which dust only affects
the observer's line-of-sight and not the BLR line-of-sights to the UV
source.

\item The  far-UV rise observed  in a few quasars, \HStre, \PGten\ and \Pks, must 
be followed by a SED turnover at higher energies.  
A detailed modeling of this flux rise, assuming dust absorption, as 
well as  the shallow  
rollover seen in \HS\ and \HE\ (and the sharper cut-off in class (D)
 HE\,1122$-$1649, Fig.\,\ref{figpks0232b}) are both consistent with the presence of 
a continuum cut-off at 18.5\,eV. Even if such a cut-off is not directly perceptible 
in individual spectra in the rest of the sample,  it is
consistent with the far-UV slope seen in the TZ02 composite.

\item The narrow continuum dip shortward of the \NEVIII\ emission line  in
the composite spectra of TZ02 and \citet{scott} is not accompanied by
a similar absorption dip blueward of \OVI\ in emission, as one might
expect if the trough was due to absorption by an outflowing ionized
wind.  Instead,  the \NEVIII\ trough could be the result  of the narrow
absorption peak that characterizes the cubic diamonds (D1) extinction
curve.
\end{list}

\acknowledgments 
This work was supported by the CONACyT grant 40096-F and the UNAM
PAPIIT grants IN113002 and IN118601. We are especially indebted to
Randal Telfer for sharing his reduced HST FOS spectra used throughout this Paper.
Diethild Starkmeth helped us with proof reading. We acknowledge the
technical support of Liliana Hern\'andez and Carmelo Guzm\'an for
configuring the Linux workstation Deneb and of Veronica Mata Acosta
and Gloria X\'ochitl P\'erez for the bibliographical research.



\clearpage 

 
\begin{figure} 
\caption{Extinction cross-sections for nanodiamonds  of radii in the range 3--25\,\AA\
from the Allende meteorite
(blue continuous line labeled A1) and for dust grains consisting of
(terrestrial) cubic diamonds (red continuous line labeled D1).  The
two long-dashed curves illustrate the effect of increasing \amax\ to
100\,\AA, corresponding to dust models A2 and D2 (labels not shown). 
Finally, the curve A3 corresponds to the case of increasing \amax\ to 200\,\AA\
(continuous orange line).  The dotted section of the terrestrial diamond curves
shortward of 413\,\AA\ corresponds to an extrapolation, as the
refraction indices are not available. The green dotted curve is the
\citet{mutschke} mass absorption coefficient curve, which is renormalized so that
its maximum coincides with the A1 curve.  The two curves barely differ,
except longward of 1200\,\AA. The black dashed curve corresponds to a
model of the ISM dust by  \citet{martin}. The ``small
grains'' dotted curve is the same model, but with \amax\ reduced (from 2500\,\AA) to
500\,\AA\ \citep{magris}.
\label{figsigmal}}
\end{figure} 

\begin{figure} 
\caption{   A qualitative description of the three main spectral classes defined in 
\S\,\ref{sec:targ}. Panel {\it a}: arbitrary scale in  $F_{\nu}$, panel {\it b}: arbitrary scale in  
$F_{\lambda}$ for the same spectra  shown in panel {\it a}. 
The $F_{\lambda}$ representation offers a clearer view of the
steepened far-UV region.  The dotted line illustrates typical
variations within a given class. Of the 106 multigrating HST
spectra available, only the 61 spectra that extended beyond the break, down to at least 900\,\AA,
could be classified reliably. The vertical dashed line represents the
position of the Lyman limit (912\,\AA\ \lar).
\label{figcartoon}}
\end{figure}

\begin{figure} 
\caption{Spectrum of PG\,1248+401 (\lar). The combined
spectrum (gratings G190H+G270H) of PG 1248+401 is shown by the
continuous thin black line and has been multiplied by the scaling
factor $0.75 \times 10^{14}$ \umm\ (hereafter, the notation \mm=0.75
will be used). Pointers indicate the position of relevant emission
lines identified by TZ02 in their composite spectrum. Red line:
absorption model as a function of rest-frame $\lambda$ using
extinction curve D1 (cubic diamond), column \NN=3.2 and a SED with
\anu=0.0. The  intrinsic quasar SED assumed (not shown) is a powerlaw
 $F_{\lambda}^{q} = (\lambda/912)^{-(2+\anu)}$, hence $F_{912}^{q}\equiv 1$ 
(in all figures). The blue line represents an
 absorption  model of equal column \NN=3.2, which uses the extinction curve A1 (Allende
 nanodiamonds).  The green line corresponds to a model with
\NN=2.8 and a dust mixture with \fdi=0.85, that is 85\% D1 grains and 15\% A1 grains.  
Underneath the green line lies the 
intergalactic model  (orange line) of \S\,\ref{sec:func} 
combined with intrinsic dust with column  \NN=1.8 and \fdi=1.0.
In all models, the continuous line part corresponds to a fiducial
spectrograph window extending from 1250 to 3600\,\AA\ (\lao), while
the dashed line represents an extension of the model into the far-UV,
down to 920\,\AA\ (\lao).  A dotted line is used in models outside
these two observer-frame windows (see \S\,\ref{sec:cate}).
\label{figpg1248}}
\end{figure}

\clearpage

\begin{figure} 
\caption{Spectrum of Pks\,0122-00  (gratings G190H+G270H) multiplied by \mm=1.1. 
The notation is the same as in Fig.\,\ref{figpg1248}.
Red line: absorption model using extinction curve D1 (cubic diamond),
column \NN=2.3 and an SED with \anu=$-0.55$. The blue line represents
a model using the extinction curve A1 (Allende
nanodiamonds) and a column \NN=2.0, while the green  line corresponds to a model with 
\NN=2.0 and a dust mixture \fdi=0.8.  Underneath the green line lies the 
intergalactic model  (orange line) of \S\,\ref{sec:func} 
combined with intrinsic dust with column  \NN=1.0 and \fdi=1.0.
\label{figpks0122}}
\end{figure} 

\begin{figure} 
\caption{Spectrum of  \PGelf\  (gratings G130H+G190H+G270H) multiplied 
by \mm=0.46. 
The notation is the same as in Fig.\,\ref{figpg1248}.  Red line:
{\it intrinsic} absorption model using extinction curve D1 (cubic
diamond) with a column \NN=1.05 and a SED with \anu=$-0.2$. The blue
line represents the same model but assuming the extinction curve A1
(Allende nanodiamonds), while the green line corresponds to a dust
mixture model with \fdi=0.6. This mixed dust model provides a satisfactory
fit of the continuum underlying the three emission lines: \OIII,
\NEVIII\ and \OIII.
\label{figpg1148a}}
\end{figure} 
 

\begin{figure} 
\caption{Histogram of the distribution  among the 44 class (A) 
spectra of the fractional contribution, \fdi, of cubic diamond dust (D1)
to the extinction curve, which results in an optimal fit of the UV break. The
contribution of the meteoritic Allende nanodiamonds (A1) is $1-\fdi$.
\label{fighistfdi}}
\end{figure} 

\begin{figure} 
\caption{Histogram of the distribution  among the 44 class (A) and 6 class (B) spectra 
of the gas columns, \NN, results in an optimal fit of the UV break.  The first bin
at \NN=0.1 corresponds to objects for which only an upper limit of
\Nh\ could be determined.
\label{fighistnh}}
\end{figure}

\begin{figure} 
\caption{Spectrum of quasar \PGten\  (gratings G150L+G270H) multiplied by \mm=0.65. 
The notation is the same as in Fig.\,\ref{figpg1248}.  Notice the
far-UV flux recovery shortward of 720\,\AA. The assumed near-UV index
is \anu=$0.13$ for all models \citep{neugebauer}.  Green  line: absorption model with  dust
 mixture  \fdi=0.3 and a column \NN=1.2. The gray   line represents
a different dust mixture of \fdi=0.6 with  same column \NN=1.2 
(it lies in the far-UV behind the green line) 
The orange line is the intergalactic model introduced in \S\,\ref{sec:func} 
to which intrinsic dust with \fdi=0.0, and column
\NN=0.50 has been added. The cyan line
is a model that assumes an SED, modified by a
shallow cut-off \Cut\ as defined in \S\,\ref{sec:cut}
(Eqn.\,\ref{eq:cut}), a dust mixture with \fdi=0.3 and a column
\NN=1.2. 
\label{figpg1008}}
\end{figure}

\begin{figure} 
\caption{Spectrum of \Pks\  (gratings G150L+G270H) multiplied 
by \mm=0.73 (black line).  
The notation is the same as in Fig.\,\ref{figpg1248}.  Notice the
far-UV flux recovery shortward of 640\,\AA. The assumed near-UV index
is \anu=$-0.4$ for all models. Red line: absorption model
of \Pks\ with column \NN=0.90 and extinction curve D1
(\fdi=1.0). The green line represents a different dust mixture with
\fdi=0.8 and column \NN=0.93.   
The  dark green line spectrum corresponds to quasar 1623.7+268B
(\mm=1.1), which also shows a rise in the far-UV.
\label{figpks0232a}}
\end{figure} 

\begin{figure} 
\caption{Spectrum of \HStre\  (gratings G190H+G270H, black line) 
multiplied by \mm=0.75. The yellow disjoint part corresponds to a GHRS spectrum with
grating G140L (same \mm).  The notation is the same as in
Fig.\,\ref{figpg1248}.  Notice the far-UV flux recovery shortward of
700\,\AA. The near-UV index is $0.0$
in all  models. Three models are shown that used the same column
\NN=1.3, but  different dust mixtures: green line: \fdi=0.6, the
purple line: \fdi=0.3, and the gray line: \fdi=0.8. 
\label{fighs1307a}}
\end{figure}

\begin{figure} 
\caption{Three class (C) spectra: cyan  line MC\,1146+111 (\mm=1.3), 
green line 3C279 (\mm=0.70) and black line 1130+106Y (\mm=0.46).  The
notation is the same as in Fig.\,\ref{figpg1248}.  Both MC\,1146+111
and 3C279 have been suitably scaled so as not to overlap with the
quasar 1130+106Y, which is being modeled.  3C279 is flat in $F_{\lambda}$, hence $\anu \simeq -2$.
Yellow line: dust absorption model of
1130+106Y assuming \anu=$-0.25$ and an extinction curve corresponding
to pure Galactic ISM extinction with \NN=9.0. 
(This last model has been multiplied by 1.3 before plotting.)
The magenta line represents
 a dust mixture of ISM-type grains (80\%) and
terrestrial diamond grains D1 (20\%). The V-band extinction implied by this model is
$A_V=0.4$\,mag.
\label{figq1130}}
\end{figure}

\begin{figure} 
\caption{Far-UV powerlaw indices \afuv\ of the quasar sample, 
as determined by TZ02, as a function of redshift. 
The solid squares (connected by a continuous line) represent an average of
\anuv\ within 5 redshift interval bins. The  long-dashed line is
the spectral index at the {\it fixed} wavelength of 800\,\AA, calculated for
the  standard intergalactic model (\nn=3.4).
\label{figalpha}}
\end{figure}

\begin{figure} 
\caption{Spectrum in $F_{\nu}$ of \HS\ (panel {\it a}, \mm=0.75) 
and \HE\ (panel {\it b}, \mm=7.8) as a function of $\lambda$. The yellow
disjoint segment corresponds to a GHRS spectrum with grating G140L (\mm=7.8).
Orange line: absorption model assuming extinction curve A3 and
{\it intergalactic} distribution of dust with \nn=3.4 (in units of
$10^{-8}$\, \cmc), \zp=0.4, \ap=+2 and \ag=$-1.5$ (see
\S\,\ref{sec:func}). The quasar energy distribution assumed is
described by the magenta dotted lines and correspond to \anu=$-0.55$
\citep{reimers89} and +1.70, for \HS\ and \HE, respectively.
The green dashed lines illustrates the effect of having {intrinsic} 
rather than intergalactic dust. The column is \NN=0.5 and the dust
composition is \fdi=0.3.  Increasing this column results in
selectively more absorption at the longer wavelength end, the opposite
of what is required.
\label{figcombi} }
\end{figure} 

\begin{figure} 
\caption{Spectrum of  \PGelf\ multiplied by \mm=0.46. 
The notation is the same as in Fig.\,\ref{figpg1248}.  The models
represent intergalactic absorption calculations, using three different
extinction curves and assuming the same SED
with \anu=$-0.2$ as in the earlier Fig.\,\ref{figpg1148a}
of the same quasar. Blue line: extinction curve A1, red line:
extinction curve D1, and orange line: extinction curve A3. The best fit
is provided by the orange line model based on the  curve A3, for
which the grain sizes extend up to
\amax=200\,\AA\ (see Fig.\,\ref{figsigmal}).  The values of \zp\ that
provide an acceptable fit to the two high-redshift quasars of
Fig.\,\ref{figcombi} are 0.8, 0.6 and 0.4 in the cases of the models
that use the curves D1, A1 and A3, respectively.  In these models,
\ap=+2 and \ag=$-1.5$, as defined in \S\,\ref{sec:func}.  
\label{figpg1148b}}
\end{figure} 
 
\begin{figure} 
\caption{Black thin line: composite energy distribution derived by
\citet{telfer} and scaled here by a factor 0.61. 
It combines both radio-loud and radio-quiet quasars. Purple line:
simulated composite absorption model assuming the dust is exclusively {\it
intergalactic} with \nn=4.7 (\S\,\ref{sec:appl}). Each synthetic
spectrum, before being absorbed and averaged, had the same SED
consisting of a powerlaw of index \anu=$-0.6$, which is represented by
the magenta dotted line.  The spectral window limits (\lao) and the
corresponding redshift of each spectrum, before being averaged, was
taken from the sample of TZ02.  The orange line is the same intergalactic
model but combined with absorption by a column, \NN=0.15, of dust
intrinsic to each quasar (\fdi=1.0). Silver dashed line: simulated composite
absorption model assuming the dust is  {\it intrinsic}  and {\it not} 
intergalactic. The column is \NN=1.0 and the mixture \fdi=0.3, both remaining
invariant with redshift.
\label{figall}}
\end{figure} 

\begin{figure} 
\caption{Transmission functions $T_{\lambda}(\zq)$ for \zq\  
values  of 0.1, 0.2, 0.5, 1.0, 1.5, 2.0, 2.8,3.5 and 5.0, assuming   the standard 
intergalactic dust model and the  extinction curve  A3.   
The parameters used to define the
function \nhf\ are  \zp=0.4, \ap=+2 and 
\ag=$-1.5$  (Eqn.\,\ref{eq:nhf} in \S\,\ref{sec:func}). 
The continuous part of each curve corresponds to the fiducial
spectrograph window of 1250--3600\,\AA\ (\lao) at the corresponding
quasar redshift. The thick gray dashed line is a
fit of the $\zq=2.8$ transmission curve using the cut-off function
\Cut\ defined in Eqn.\,\ref{eq:cut}, with \labr=670\,\AA,
\ad=$-1.6$ and $f=2.8$.
\label{figtrans}}
\end{figure} 

\begin{figure} 
\caption{A repetition of the spectrum of \HStre\  of Fig.\,\ref{fighs1307a}.  
In all models, the assumed near-UV index is $0.0$, as in
Fig.\,\ref{fighs1307a}. The  brown dashed line corresponds to the
standard intergalactic model introduced in \S\,\ref{sec:func}.  The
orange line is the same intergalactic model, but combined with an
{\it intrinsic} dust column of \NN=0.8 with \fdi=0.6.  The green dashed line shows
the contribution by intrinsic dust in the case of this combined
intergalactic model (orange line).  The cyan line is the intrinsic
dust model, but assuming that the powerlaw is modified by a
shallow cut-off, as described in \S\,\ref{sec:cut} (Eqn.\,\ref{eq:cut}
with \labr=670\,\AA, $f=2.8$ and $\ad=-1.6$). The column is \NN=1.0 and the
dust mixture \fdi=0.6.
\label{fighs1307b}}
\end{figure}  
  
\begin{figure} 
\caption{A repetition of the spectrum of \Pks\ of Fig.\,\ref{figpks0232a}.  
In all models, the assumed near-UV index is \anu=$-0.2$, as in
Fig.\,\ref{figpks0232a}.  The orange line corresponds to the standard
intergalactic model introduced in \S\,\ref{sec:gala}, but combined
with an intrinsic dust column of \NN=0.18 with mixture \fdi=1.0.  The
cyan line (underneath the orange line) 
is the intrinsic dust model, but assuming that the powerlaw
is modified by a shallow cut-off, as described in \S\,\ref{sec:cut}
(Eqn.\,\ref{eq:cut} with \labr=670\,\AA, $f=2.8$ and $\ad=-1.6$). The
column is \NN=0.62 and the dust mixture \fdi=1.0 The blue line
spectrum at the bottom corresponds to the class (D) spectrum of
HE\,1122$-$1649 (\mm=0.44). The gray and magenta lines correspond to
a powerlaw with $\anu=-0.6$, multiplied by the cut-off function \Cut\
of Eqn.\,\ref{eq:cut}, assuming a form factor $f$ of 2.8 and 10,
respectively. In both cases, the break wavelength is \labr=670\,\AA\
and the index change $\ad=-2.0$.
\label{figpks0232b}}
\end{figure} 
 
\begin{figure} 
\caption{Black thin line: composite energy distribution.  Black dotted line:  an
SED consisting of a powerlaw of index \anu=$-0.6$
multiplied by the cut-off function \Cut\ described by
Eqn.\,\ref{eq:cut}.  The parameters are the same as those inferred
from the $z=2.8$ transmission curve in Fig.\,\ref{figtrans}, that is
\labr=670\,\AA, \ad=$-1.6$ and $f$=2.8.  The darker gray line is a composite
simulation, assuming intrinsic dust and the above powerlaw modified by the function
\Cut. The dust column is \NN=0.8 with a mixture \fdi=0.3.  Silver
dashed line: the  composite simulation of previous Fig.\,\ref{figall},
assuming intrinsic dust, but {\it without} an intrinsic cut-off (pure powerlaw).
\label{figallb}}
\end{figure}

\clearpage  



\begin{thebibliography}{} 
 
\bibitem[Acke \& van den Ancker(2004)]{acke04} Acke, B., \& 
van den Ancker, M.~E.\ 2004, \aap, 426, 151 

\bibitem[Baldry \& Glazebrook(2003)]{baldry} Baldry, I.~K., \& 
Glazebrook, K.\ 2003, \apj, 593, 258 


\bibitem[Barstow et al.(2003)]{barstow} Barstow, M.~A., et al.\ 
2003, \procspie, 4854, 364 

\bibitem[Binette et al.(1988)]{binette88} Binette, L., Robinson, 
A., \& Courvoisier, T.~J.-L.\ 1988, \aap, 194, 65 

\bibitem[Binette, Magris \& Martin(1993)]{magris} Binette, 
L., Magris, G., \& Martin, P.\ 1993, in proc. of First Light in the Universe. Stars or 
QSO's?, eds. B. Rocca-Volmerange, B. Guiderdoni, M. Dennefeld and J. Tran Thanh Van, 
(Paris: Editions Fronti\`eres), p. 243 

\bibitem[Binette et al.(2005a)]{bi05a} Binette, L., Morisset, C., \& Haro-Corzo, S., 
in proc. of {\it The ninth
Texas-Mexico Conference on Astrophysics},  April
13--16, 2005, in San Antonio, Texas, eds. S. Torres \& G. MacAlpine,  \rmxaa\ (Conf. Ser.), in press

\bibitem[Binette et al.(2005b)]{bi05b} Binette, L., 
Krongold, Y. , Magris C., G., \& de Diego, J. A. 2005b, in proc. of "Triggering relativistic jets" 
held in Cozumel, 28 March  -- 1 April 2005, Eds. W. Lee and E. Ramirez-Ruiz, \rmxaa\ (Conf. Ser.), 
in press

\bibitem[Binette et al.(2005c)]{bi05c} Binette, L., Mutschke, H., \& Andersen, A. C., 2005c,  
Astronomishe Nachrichten, in Proc. of 
``Granada Workshop on High Redshift Radio Galaxies'', Granada, 
18--20 April 2005, Ed. M. Villar-Martin et al., Astronomische Nachrichten, in press 



\bibitem[Binette et al.(2003)]{binc} Binette, L.,  Rodr\'\i
guez-Mart\'\i nez, M., Haro-Corzo, S., \& Ballinas, I. 2003,  \apj, 590, 58

\bibitem[Bohren \& Huffman(1983)]{bohren} Bohren, C. F., \& Huffman, D.R. 1983,
Absorption by Small Particles (New York: Wiley)


  

\bibitem[Calura \& Matteucci(2004a)]{calura04a} Calura, F., \& 
Matteucci, F.\ 200ab, \mnras, 350, 351 
 
\bibitem[Calura \& Matteucci(2004b)]{calura04b}  Calura, F., \& Matteucci, F. 2004b, 
\mnras, 351, 384 

\bibitem[Daulton et al.(1995)]{daulton95} Daulton, T.~L., Ozima, 
M., \& Shukolyukov, Y.\ 1995, Lunar and Planetary Institute Conference 
Abstracts, 26, 313 

\bibitem[Duley \& Grishko(2001)]{duley01} Duley, W.~W., \& 
Grishko, V.~I.\ 2001, \apjl, 554, L209 

\bibitem[Edwards \& Philipp(1985)]{edwards} Edwards, D. F., \& Philipp, H. R. 1985,
in Handbook of Optical Constants of Solids, ed. E. D. Palik (Orlando:
Academic Press), 665
 
\bibitem[Ferland et al.(1996)]{ferland} Ferland, G.~J., 
Baldwin, J.~A., Korista, K.~T., Hamann, F., Carswell, R.~F., Phillips, M., 
Wilkes, B., \& Williams, R.~E.\ 1996, \apj, 461, 683 

\bibitem[Ferrara et al.(1991)]{ferrara91} Ferrara, A., Ferrini, 
F., Barsella, B., \& Franco, J.\ 1991, \apj, 381, 137 

 


\bibitem[Guillois et al.(1999)]{guillois99} Guillois, O., Ledoux, 
G., \& Reynaud, C.\ 1999, \apjl, 521, L133 


\bibitem[Haro-Corzo et al.(2005)]{sinhue} Haro-Corzo, S., Binette, L., 
Benitez, E., Krongold, Y. 2005, in preparation

  
\bibitem[Jones et al.(2004a)]{jones04a} Jones, A.~P., 
d'Hendecourt, L.~B., Sheu, S.-Y., Chang, H.-C., Cheng, C.-L., \& Hill, 
H.~G.~M.\ 2004, \aap, 416, 235 

\bibitem[Jones et al.(2004b)]{jones04b} Jones, A.~P., 
d'Hendecourt,  L.B.\ 2004, in ASP Conf. Series, Astrophysics of Dust, V. 309, 589

\bibitem[Koratkar \& Blaes(1999)]{koratkar99} Koratkar, A., \& 
Blaes, O.\ 1999, \pasp, 111, 1 

\bibitem[Korista, Ferland \& Baldwin(1997)]{korista} Korista, K.,  
Ferland, G., \& Baldwin, J. 1997, \apj, 487, 555


\bibitem[Kouchi et al.(2005)]{kouchi05} Kouchi, A., Nakano, H., Kimura, Y., 
\& Kaito, C.\ 2005,   \apjl, submitted

\bibitem[Kriss et al.(2001)]{kriss}
Kriss, G. A., et al. 2001, Science, 293, 1112
 
\bibitem[Laor \& Draine(1993)]{laor93} Laor, A., \& Draine, 
B.~T.\ 1993, \apj, 402, 441 


\bibitem[Lewis et al.(1987)]{lewis87} Lewis, R.~S., Ming, T., 
Wacker, J.~F., Anders, E., \& Steel, E.\ 1987, \nat, 326, 160 

\bibitem[Lewis et al.(1989)]{lewis89} Lewis, R.~S., Anders, E., 
\& Draine, B.~T.\ 1989, \nat, 339, 117 

\bibitem[Martin \& Rouleau(1991)]{martin} Martin, P.~G., \& 
Rouleau, F.\ 1991, Extreme Ultraviolet Astronomy, ed. R.F. Malina and
S. Bowyer (Pergamon, Oxford), p. 341

\bibitem[Mathews \& Ferland(1987)]{mathews} Mathews, W.~G., \& 
Ferland, G.~J.\ 1987, \apj, 323, 456 

 
\bibitem[Mutschke et al.(2004)]{mutschke} Mutschke, H., 
Andersen, A.~C., J{\" a}ger, C., Henning, T., \& Braatz, A.\ 2004, \aap,
423, 983 


\bibitem[Neugebauer et al.(1987)]{neugebauer} Neugebauer, G., 
Green, R.~F., Matthews, K., Schmidt, M., Soifer, B.~T., \& Bennett, J.\ 
1987, \apjs, 63, 615 

\bibitem[Nuth \& Allen(1992)]{nuth92} Nuth, J.~A., \& Allen, 
J.~E.\ 1992, \apss, 196, 117 

\bibitem[O'Brien et al.(1988)]{obrien} O'Brien, P.~T., Wilson, 
R., \& Gondhalekar, P.~M.\ 1988, \mnras, 233, 801 

\bibitem[Perlmutter et al.(1999)]{perl99} Perlmutter, S., et 
al.\ 1999, \apj, 517, 565 
 
\bibitem[Reimers et al.(1989)]{reimers89} Reimers, D., Clavel, 
J., Groote, D., Engels, D., Hagen, H.~J., Naylor, T., Wamsteker, W., \& 
Hopp, U.\ 1989, \aap, 218, 71 


\bibitem[Rouan et al.(2004)]{rouan04} Rouan, D., et al.\ 2004, 
\aap, 417, L1 
 

\bibitem[Scott et al.(2004)]{scott} Scott, J., Kriss, G. A., Brotherton, 
M. S., Green, R. F.,  Hutchings, J., Shull, J. M., \& Zheng, W. 2004, \apj, 615, 135

\bibitem[Shang et al.(2004)]{shang} Shang, Z., Brotherton, M. S., Green, R. F., Kriss, G. A., Scott, J., Quijano, J. K., Blaes, O., Hubeny, I., Hutchings, J., Kaiser, M. E., Koratkar, A., Oegerle, W., \& Zheng, W. 2005, \apj,  619, 41

\bibitem[Spergel et al.(2003)]{spergel} Spergel, D.~N., et al.\ 
2003, \apjs, 148, 175 

\bibitem[Telfer et al.(2002)]{telfer} Telfer, R. C., Zheng, W., Kriss, G. A., 
 \& Davidsen, A. F. 2002, \apj, 565, 773 (TZ02)
 
\bibitem[Tielens et al.(1987)]{tielens87} Tielens, A.~G.~G.~M., 
Seab, C.~G., Hollenbach, D.~J., \& McKee, C.~F.\ 1987, \apjl, 319, L109 

\bibitem[Van Kerckhoven et al.(2002)]{van02} Van Kerckhoven, 
C., Tielens, A.~G.~G.~M., \& Waelkens, C.\ 2002, \aap, 384, 568 


\bibitem[Whittet(2002)]{whittet} Whittet, D. C. B. 2002, 
Dust in the galactic environment,  Second edition, (Bristol:IOP) 


\bibitem[Zheng et al.(1997)]{zheng} Zheng, W., Kriss, G. A., Telfer, R. C., 
Grimes, J. P., \& Davidsen, A. F. 1997, \apj, 475, 469 


\end{thebibliography}
\end{document}